\documentclass[]{spie}  

\usepackage{amsmath,amsfonts,amssymb}
\usepackage{graphicx}
\usepackage[colorlinks=true, allcolors=blue]{hyperref}
\usepackage[colorinlistoftodos]{todonotes}
\usepackage{multirow}
\usepackage{mathrsfs}
\usepackage{xspace}

\newcommand{\tbath}{$T_{\mbox{\scriptsize bath}}$\xspace}

\newcommand{\apj}{    {\it ApJ}}

\newcommand{\mnras}{  {\it MNRAS}}
\newcommand{\jcap}{  {\it JCAP}}

\newcommand{\procspie}{   {\it Proceedings of SPIE}}

\newcommand{\prd}{    {\it Phys. Rev. D}}

\title{Studies of Systematic Uncertainties for Simons Observatory: Detector Array Effects}

\author[a]{Kevin T. Crowley}
\author[b]{Sara M. Simon}
\author[c]{Max Silva-Feaver}
\author[d]{Neil Goeckner-Wald}
\author[d]{Aamir Ali}
\author[h]{Jason Austermann}
\author[n]{Michael L. Brown}
\author[d]{Yuji Chinone}
\author[d]{Ari Cukierman}
\author[h]{Bradley Dober}
\author[h]{Shannon M. Duff}
\author[a]{Jo Dunkley}
\author[f]{Josquin Errard}
\author[m]{Giulio Fabbian}
\author[i]{Patricio A. Gallardo}
\author[a]{Shuay-Pwu Patty Ho}
\author[h]{Johannes Hubmayr}
\author[c]{Brian Keating}
\author[e]{Akito Kusaka}
\author[n]{Nialh McCallum}
\author[b]{Jeff McMahon}
\author[g]{Federico Nati}
\author[i]{Michael D. Niemack}
\author[k]{Giuseppe Puglisi}
\author[e]{Mayuri Sathyanarayana Rao}
\author[j]{Christian L. Reichardt}
\author[f]{Maria Salatino}
\author[c]{Praween Siritanasak}
\author[a]{Suzanne Staggs}
\author[e]{Aritoki Suzuki}
\author[c]{Grant Teply}
\author[n]{Daniel B. Thomas}
\author[h]{Joel N. Ullom}
\author[f]{Clara Verg\`{e}s}
\author[h]{Michael R. Vissers}
\author[d]{Benjamin Westbrook}
\author[l]{Edward J. Wollack}
\author[g]{Zhilei Xu}
\author[g]{Ningfeng Zhu}
\affil[a]{Department of Physics, Princeton University, Princeton, NJ, USA}
\affil[b]{Department of Physics, University of Michigan, Ann Arbor, MI, USA}
\affil[c]{Department of Physics, University of California San Diego, La Jolla, CA, USA}
\affil[d]{Department of Physics, University of California Berkeley, Berkeley, CA, USA}
\affil[e]{Physics Division, Lawrence Berkeley National Laboratory, Berkeley, CA, USA}
\affil[f]{AstroParticule et Cosmologie, Univ Paris Diderot, CNRS/IN2P3,CEA/Irfu, Obs de Paris, Sorbonne Paris Cité, France}
\affil[g]{Department of Physics and Astronomy, University of Pennsylvania, Philadelphia, PA, USA}
\affil[h]{Quantum Sensors Group, NIST, Boulder, CO, USA}
\affil[i]{Department of Physics, Cornell University, Ithaca, NY, USA}
\affil[j]{School of Physics, University of Melbourne, Melbourne, Australia}
\affil[k]{Department of Physics, Stanford University, Stanford, CA, USA}
\affil[l]{NASA/Goddard Space Flight Center, Greenbelt, MD, USA}
\affil[m]{Institut d’Astrophysique Spatiale, CNRS (UMR 8617), Univ. Paris-Sud, Université Paris-Saclay, B$\mathrm{\hat{a}}$t.121, 91405 Orsay, France}
\affil[n]{Jodrell Bank Centre for Astrophysics, University of Manchester, Manchester, UK}

\authorinfo{Send correspondence to K.T.C. and S.M.S.\\K.T.C.: E-mail: ktc2@princeton.edu\\S.M.S.: E-mail: smsimon@umich.edu}

\begin{document}
\maketitle

\begin{abstract}
In this proceeding, we present studies of instrumental systematic effects for the Simons Obsevatory (SO) that are associated with the detector system and its interaction with the full SO experimental systems. SO will measure the Cosmic Microwave Background (CMB) temperature and polarization anisotropies over a wide range of angular scales in six bands with bandcenters spanning from 27 GHz to 270 GHz. We explore effects including intensity-to-polarization leakage due to coupling optics, bolometer nonlinearity, uncalibrated gain variations of bolometers, and readout crosstalk. We model the level of signal contamination, discuss proposed mitigation schemes, and present instrument requirements to inform the design of SO and future CMB projects.
\end{abstract}

\section{Introduction}
The cosmic microwave background (CMB) contains a wealth of information in both its temperature and polarization, including information that can help determine the nature of dark energy and dark matter, the mass and number of neutrino species, and if there was a period of inflation shortly after the universe began. The linear polarization of the CMB can be decomposed into even (E-mode) and odd (B-mode) parity components. If inflation occurred, the gravitational waves would have produced both E-modes and B-modes. Since E-modes are also produced by density perturbations, the detection of a B-mode signal would be particularly powerful in giving additional support to inflationary models~\cite{Seljak_1997, Kami_1997}. The amplitude of this primordial B-mode signal is expected to peak on degree-angular scales (multipole moments of $\ell \sim 100$), and its amplitude is quantified by the tensor-to-scalar ratio $r$. There is also a contribution to B-modes at sub-degree scales from the gravitational lensing of E-modes into B-modes from intervening large-scale structure that, in combination with E-modes and temperature anisotropies, holds information about the mass and number of neutrinos and the nature of dark energy and dark matter~\cite{lensing}. However, B-modes are faint, and the primordial B-mode signal has yet to be measured. This has pushed CMB science toward increasingly sensitive experiments where calibration and the mitigation of systematic effects are critical. Further complicating measurements is the contamination of CMB polarization by foregrounds from synchrotron and dust emission. These foreground signals have a different frequency dependence than the CMB polarization, so in principle, these signatures can be removed at the required fidelity if appropriately characterized as a function of frequency.

  The Simons Observatory (SO) will observe the CMB in both temperature and polarization over a wide range of frequencies (27-270~GHz) and angular scales. SO will field a $\sim$6~m crossed-Dragone, large-aperture telescope (LAT) for observations at small angular scales and several small-aperture ($\sim$0.5~m) telescopes (SATs) for large angular scale measurements. SO plans to push to high sensitivity by deploying $\sim$50 multichroic detector arrays in its initial configuration. This represents more millimeter-wave detectors for the observation of the CMB than have yet been deployed elsewhere and represents a critical step toward next-generation experiments like CMB-S4~\cite{S4}. To fully utilize this sensitivity, the systematic effects must be well-characterized and mitigated through the instrument design. In this paper, we focus on systematic effects originating from the detector system, which include the optical coupling (feedhorns and lenslets), detectors, and signal readout. This paper is part of a series of papers on the systematic and calibration studies for SO~\cite{salatino18,bryan18,gallardo18}. We are combining the detailed results of the full SO systematics and calibration studies into a comprehensive study that will be released in a series of future papers to the community for use in developing future CMB experiments such as CMB-S4. Here we take an in-depth look at several of the most important detector array systematics, and note that further information on detector array systematic effects that are not included in this paper will be included in the full systematics study papers. In Sec.~\ref{sec:analysis_frameworks}, we introduce two analysis frameworks used to model detector systematic effects: a time-domain systematics pipeline and a map-based systematics pipeline. Section~\ref{sec:IP_horn_lenslet} discuses simulations of polarization leakage from feedhorns and lenslets. In Sec.~\ref{sec:nonlinearity} and~\ref{sec:gain_drift}, we discuss the possible spurious polarization induced by two sources of time-varying differential gain between polarization-pair detectors: long-timescale gain drifts and bath temperature fluctuations, respectively. Finally, in Sec.~\ref{sec:crosstalk}, we discuss crosstalk studies for frequency-division readout schemes.
\section{Analysis Frameworks} \label{sec:analysis_frameworks}
We use two analysis frameworks for our systematic studies. The first is a map-based systematics pipeline that is used for beam-related effects. In this pipeline, systematics are modeled and incorporated in the map domain. We use this framework for studies of the polarization leakage arising from the detector array optical coupling. The second is a time-domain systematics pipeline where systematic effects are modeled and introduced into the data channel timestreams. This pipeline is used for non-linearity, gain drift, and crosstalk studies.
\subsection{Map-Based Systematics Pipeline}
The map-based simulation pipeline generates realizations of the CMB sky, instrument noise, and the combination of both for a given fraction of the sky $f_{sky}$ and instrument noise realization. The simulator then convolves the sky simulations with instrument beams, uses the pseudo-C$_\ell$ method, and calculates 128 realizations of noise, signal, and signal with noise~\cite{Hivon_2002,Smith_2006}. Using these transfer functions, individual realizations of noise and the CMB are analyzed to estimate the expected signal for both a perfectly Gaussian beam (no systematic effects) and simulations of the beams with systematic effects included. The $\Delta\chi^2$ between these two cases is then calculated for a given $\ell$ range and can be averaged over a given number of individual realizations. In the analysis presented in Sec.~\ref{sec:lens_feed_sac} for the SAT, we take $f_{sky}=0.12$ and a white noise level of 2~$\mathrm{\mu}$K-arcminute, which is the goal sensitivity for SO. The beams with systematics are the simulated T$\rightarrow$Q and U$\rightarrow$Q beams calculated in Sec.~\ref{sec:pol_beam}. We take the flat sky approximation, which is valid in the SAT case.

\subsection{Time-Domain Systematics Pipeline}
\label{subsec:s4cmb}
To study the effects of detector-level systematics, which often affect the time-domain signals of the detectors in an array, we use the public Python code \textit{s4cmb} \footnote{J. Peloton, https://github.com/JulienPeloton/s4cmb/}. This code is derived from a systematics pipeline built for data analysis for \textsc{polarbear}~\cite{PB_2017}. 

Each simulation begins by producing an input sky map. We generate maps based on a set of fiducial bandpowers which include CMB lensing and draw a sky realization using HealPy \footnote{HealPy, https://github.com/healpy/healpy}, a Python wrapper for the HEALPix library \footnote{http://healpix.sf.net}. The focal plane instrumentation is represented schematically as $N$ pairs of detectors, where each pair is assumed to have two detectors of exactly orthogonal polarization sensitivity, read out through frequency-domain multiplexing electronics~\cite{Dfmux_readout_2012,Mates_Thesis_2011}. The readout architecture is only relevant in defining groups of detectors that share readout components. The physical size of the focal plane and its projected size on the sky define the optical parameters of the focal plane.

Once an instrument description is defined, we initialize a scanning strategy assuming either shallow(wide)-field or deep(small)-field observations. Together, the scan strategy and instrument descriptions define a map-to-time domain projection matrix. This is used to scan the sky according to each detector's pointing. Each constant elevation scan (CES) is about four hours, and each scan strategy features a set number of total CESes, with only one observed per day. This mimics a $\sim$ 20\% observing efficiency and is a hard-coded element of the basic scanning strategies available in the software. This efficiency does not reflect the  expectations for SO observing, where we expect the efficiency to be significantly higher.

After projection into the time domain, white noise or a sum of white noise and correlated noise is added to detector timestreams. The white noise parameters are estimated from the expected SO array sensitivities expressed in noise-equivalent temperature (NET)~\cite{hill18} distributed evenly among $N$ simulated detectors as:
\begin{equation}
NET_{\mbox{\scriptsize det}} = NET_{\mbox{\scriptsize array}} \sqrt{N_{\mbox{\scriptsize det}}}\,\,.
\label{eq:perdet_sensitivity}
\end{equation}
The correlated noise is parameterized as $1/f^{a}$, where $f$ is the frequency of the noise fluctuations and $a>0$. In these simulations, this noise is completely correlated across subsets of detectors in the array. This simulates atmospheric modes that span significant scales on the sky. We define the scale of these atmospheric fluctuations according to some number $n_{\mbox{\scriptsize cloud}}$ of correlated array regions. The correlated noise also has a correlation length of $t_{\mbox{\scriptsize corr}}$, which we set at five minutes in our initial simulations.

Finally, time-domain systematic effects are applied to the per-detector data. The timestreams with noise and systematic effects included are projected into an output map. To quantify the impact of the modeled systematic errors, we can use the output map directly or compute the power spectra of the output map.

\section{Detector Array Optical Coupling}\label{sec:IP_horn_lenslet}
SO plans to field both feedhorn-coupled detectors and lenslet-coupled detectors. Light from the spline-profiled feedhorns is coupled to the detectors with an orthomode transducer (OMT), and modeling the feedhorn gives the beam response of the system.  To determine the full beam response of the detectors on the lenslet-coupled sinuous antenna array, the full lenslet and sinuous antenna structure must be modeled together. To model the detector array beams, we use High Frequency Structure Simulator (HFSS)\footnote{ANSYS, Inc. Canonsburg, PA 15317}.

Beam asymmetries from the optical coupling to the detector array can cause leakage from temperature to polarization (T$\rightarrow$P) and E-modes to B-modes (E$\rightarrow$B)~\cite{Shimon_2008}. The simplest way to model this leakage is by assuming that the polarized signal is obtained by differencing a pair of detectors at the same frequency on each pixel that are sensitive to orthogonal polarizations. In a pair-differenced system, differential beam effects from the two orthogonal detector beams and thus the leakage are maximized. However, accounting for beam asymmetries in the analysis can mitigate the leakage by an order of magnitude or more depending on how well the telescope beams are characterized by planets, point sources, and external calibration sources. Further, many CMB analyses like maximum-likelihood map-makers do not rely on explicit pair-differencing to recover the polarized signal~\cite{Naess_2014}. In the presence of a continuously-rotating half-wave plate (CRHWP) as in the SATs, differential beam effects are mitigated because each detector independently measures $I$, $Q$, and $U$. To estimate the leakage with a CRHWP, one would have to propagate the beam analysis through a full time-domain analysis that includes demodulating the signal from the CRHWP rotation~\cite{ABS_2014}. In all cases, the pair-differenced scheme represents an upper limit on the total polarization leakage, so we employ this method to rapidly check that the polarization leakage from the optical coupling designs are at an acceptable level to achieve the SO science goals. This estimation can also be used within SO to contribute to critical design decisions like the optimal pixel size and selecting the final feedhorn and lenslet designs. We note that the final designs of feedhorns and lenslets for SO are not complete, so we use preliminary designs for the 90/150~GHz multichroic detector bands in this work.

\subsection{Polarization Leakage Estimation} \label{sec:pol_beam}
The polarization leakages in the power spectra are estimated using the simulated co- and cross-polar beams modeled in HFSS. First the leakage beams are determined following the analysis in Simon, 2016~\cite{Simon_Thesis_2016,Simon_SPIE_2016}. Assuming a pair of detectors sensitive to orthogonal polarizations in one aperture, the electric fields on the sky $E_x$ and $E_y$ are coupled to the electric field in the detectors $E_a$ and $E_b$ by
\begin{equation}\label{eqn:beams}
  \begin{bmatrix}
  E_a\\
  E_b
  \end{bmatrix}
  =
  \begin{bmatrix}
  \beta_{ax} & \beta_{ay}\\
  \beta_{bx} & \beta_{by}
  \end{bmatrix}
  \begin{bmatrix}
  E_x\\
  E_y
  \end{bmatrix}
  \,\,\, ,
\end{equation}
where $a$ and $x$ as well as $b$ and $y$ are aligned along the boresight. Here $\beta_{ax}$ and $\beta_{by}$ are the complex 2D co-polar beams, and $\beta_{ay}$ and $\beta_{bx}$ are the complex 2D cross-polar beams. To calculate the far field beams, the co- and cross-polar beams from HFSS are first masked such that they go to zero outside of the Lyot stop ($\sim 17^{\circ}$ for the SAT and $\sim 13^{\circ}$ for the LAT), the beams are corrected for any defocus from the HFSS simulations, and a 2D Fourier transform is performed.

For an ideal detector pair, the measured 2D polarized signal $P$ would be
\begin{equation}\label{eqn:p1}
P=|E_a|^2-|E_b|^2 \,\,\,\,.
\end{equation}

Substituting Eq.~\ref{eqn:beams} into Eq.~\ref{eqn:p1} and expressing the result in terms of the Stokes parameters gives
\begin{equation}
P=\sigma I + \delta Q + \epsilon U+ \gamma V\,\,\,\, ,
\end{equation}
where the coefficients are the beam couplings from $I$, $Q$, $U$, and $V$ into $P$. The beam couplings are then given by
\begin{align}\label{eqn:leakage beams}
\sigma & =  \frac{1}{2} (|\beta_{ax}|^2 + |\beta_{ay}|^2 - |\beta_{bx}|^2 - |\beta_{by}|^2) \\
\delta & =   \frac{1}{2} (|\beta_{ax}|^2 - |\beta_{ay}|^2 - |\beta_{bx}|^2 + |\beta_{by}|^2) \\
\epsilon & =  \mathrm{Re}(\beta_{ax}^{*} \beta_{ay} - \beta_{bx}\beta_{by}^{*} )  \\
\gamma & =  -\mathrm{Im}(\beta_{ax} \beta_{ay}^{*} + \beta_{bx}^{*}\beta_{by} )\,\,\,\,\,.
\end{align}
For an ideal optical system including the detector and telescope measuring $Q$, $\delta$ is a Gaussian beam with a peak of one and $\sigma=\epsilon=\gamma=0$. The beams are normalized by the maximum of $\delta$ and averaged across each observation band~\cite{Simon_SPIE_2016}.

We estimate the impact of this leakage on the power spectra from these band-averaged beams with two methods: a map-based systematics pipeline and a window function method. The map-based method gives a full estimation of the polarization leakage using the leakage beams and simulated maps and realistic noise estimates. This method is useful for determining how much of the temperature to polarization leakage goes into E-modes versus B-modes. With the window function method, the leakage is estimated using the calculated window functions of the signal and leakage beams. This simplified method is quick to model, but only gives total polarization leakage. However, it is still good as a check on the worst-case scenario. 

In the window function method, for each beam, the magnitude squared of the Fourier transform of the averaged far field beams is calculated and normalized by the maximum of the transformed $\delta$ beam. Next the 2D functions are binned radially to make a 1D window function. To account for the rest of the telescope optics, we scale the ordinate according to the size of the output aperture to create an $\ell$-space window function. The measured spectra are then estimated by multiplying simulations of the EE and BB polarization spectra by the $\delta$ window function, the temperature to polarization leakage spectrum is determined by multiplying the simulated TT spectrum by the $\sigma$ window function, and the EE to BB leakage is determined by multiplying the simulated EE spectrum by the $\epsilon$ window function. 

\subsection{Variation in Systematics with Pixel Size}

\begin{figure}[h!]
\centering
\includegraphics[width=\textwidth]{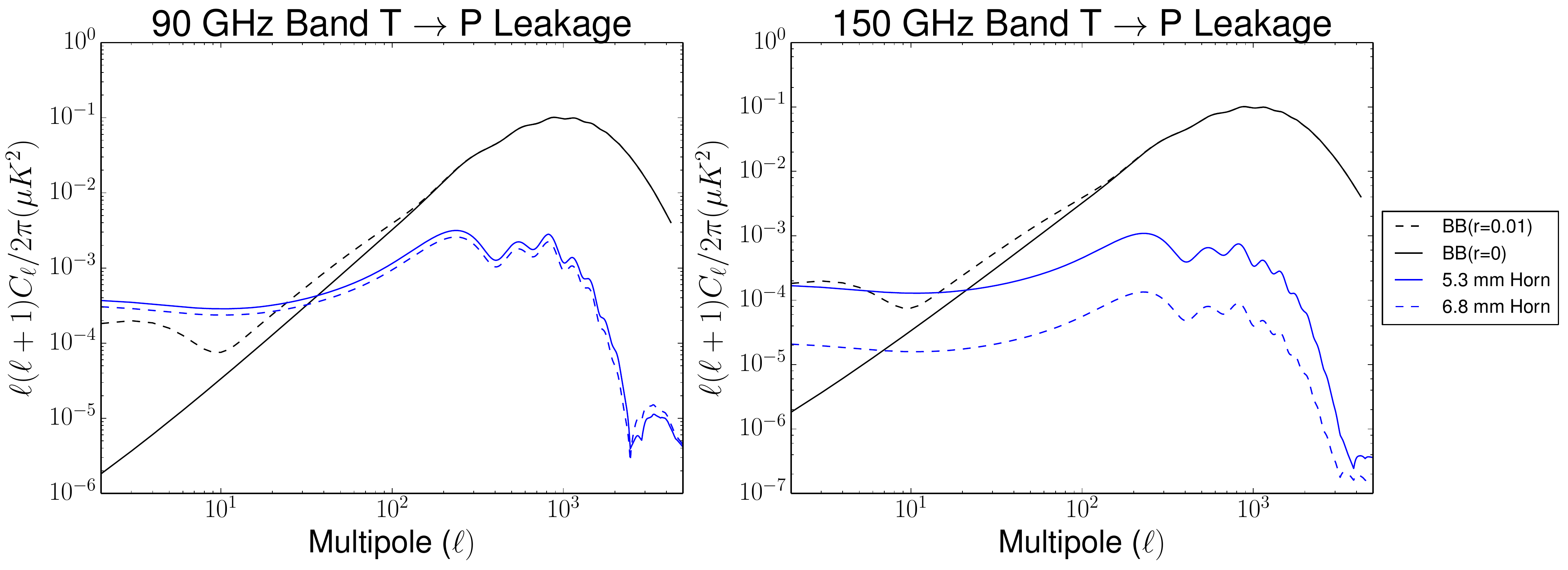}
\caption{The temperature to polarization leakage with no mitigation from a feedhorn design for a 5.3~mm pixel size (solid) and a 6.8~mm pixel size (dashed) estimated with the window function method for the SO LAT for the 90~GHz (left) and 150~GHz (right) bands is shown above in blue. While the feedhorn T$\rightarrow$P leakage primarily goes into E-modes, the full T$\rightarrow$P leakage is plotted with simulated B-modes in black for $r=0$ (solid) and $r=0.1$ (dashed). Even if all of the T$\rightarrow$P leakage went into the B-mode signal, it would be negligible on the LAT, especially once the suppression from beam calibration is included. As pixel size increases, the total T$\rightarrow$P leakage of the feedhorns decreases. As expected from the waveguide cutoff, the improvement in the lower band is smaller than that in the upper band.}
\label{fig:MF_horn_TP_leakage_pixel_size}
\end{figure}

The feedhorns for SO are designed through a Markov chain Monte Carlo (MCMC) optimization between beam coupling efficiency and beam symmetry~\cite{Simon_SPIE_2016,Simon_Thesis_2016}. As pixel size increases, the feedhorn aperture increases, which yields both increased beam coupling efficiency and beam symmetry. Increased beam symmetry results in lower T$\rightarrow$P leakage in both frequency bands, though the improvement in the lower band is usually smaller because the waveguide cutoff of the feedhorn can cause beam distortion. This trend is illustrated in Fig.~\ref{fig:MF_horn_TP_leakage_pixel_size}, which shows the total T$\rightarrow$P leakage estimated with the window function method for feedhorns designed for a 5.3~mm and 6.8~mm pixel size on the LAT. We note that simulations using the window function method for the LAT have artificially inflated leakage below $\ell \lesssim 100$ due to the simulation resolution of the window function.

\begin{figure}[h!]
\centering
\includegraphics[width=\textwidth]{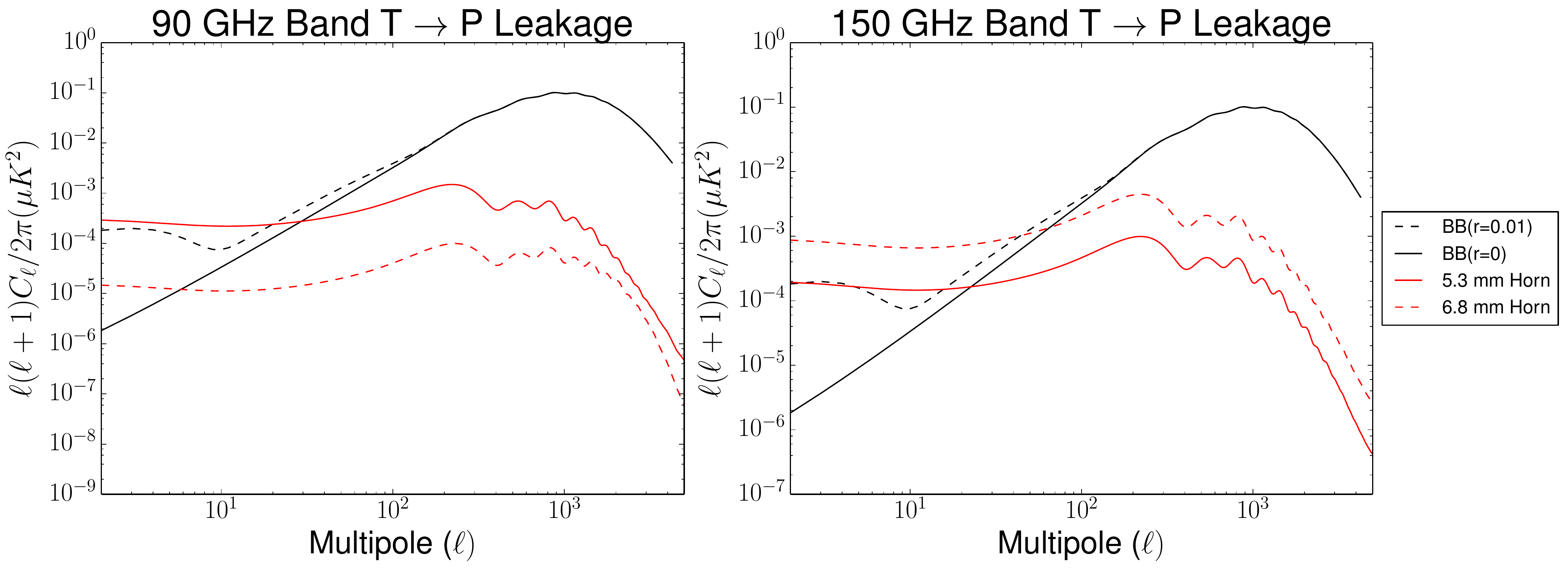}
\caption{The temperature to polarization leakage with no mitigation from the lenslet and sinuous antenna design for a 5.3~mm pixel size (solid) and a 6.8~mm pixel size (dashed) estimated with the window function method for the SO LAT for the 90~GHz (left) and 150~GHz (right) bands is shown above in red with the same convention as Fig.~\ref{fig:MF_horn_TP_leakage_pixel_size}. As pixel size increases, the T$\rightarrow$P leakage decreases in the low band and increases in the high band. We note that even if all of the T$\rightarrow$P leakage went into the B-mode signal, it would be negligible on the LAT, especially once the suppression from beam calibration is included.}
\label{fig:MF_lenslet_TP_leakage_pixel_size}
\end{figure}

For the sinuous antenna and lenslet architecture, as pixel size increases, there is no overall decrease in the level of T$\rightarrow$P leakage. Instead, as the pixel size increases, the level of leakage in the high band increases, while the level of the leakage decreases in the low band and vice versa for smaller pixel sizes. Figure~\ref{fig:MF_lenslet_TP_leakage_pixel_size} shows the trend in T$\rightarrow$P leakage with pixel size for the lenslet and sinuous antenna architecture at 5.3~mm and 6.8~mm for the LAT. Systematic analyses along with sensitivity and layout constraints were used to determine the final pixel size for the 90/150~GHz arrays: 5.3~mm for the horn-coupled pixels, and 5.6~mm for the lenslet and sinuous antenna architecture.

\subsection{Comparison Between Architectures}\label{sec:lens_feed_sac}

In addition to the differences in T$\rightarrow$P leakage with pixel size, there are a few differences in the systematic performance between the feedhorn-coupled OMT antennae and lenslet-coupled sinuous antennae. Feedhorns are tunable between beam coupling efficiency and beam symmetry in their design. The relative weighting between these two properties is set by the penalty function used for the MCMC optimization, so the level of leakage can be decreased at the cost of decreasing coupling efficiency and vice versa.

\begin{figure}[h!]
\centering
\includegraphics[width=\textwidth]{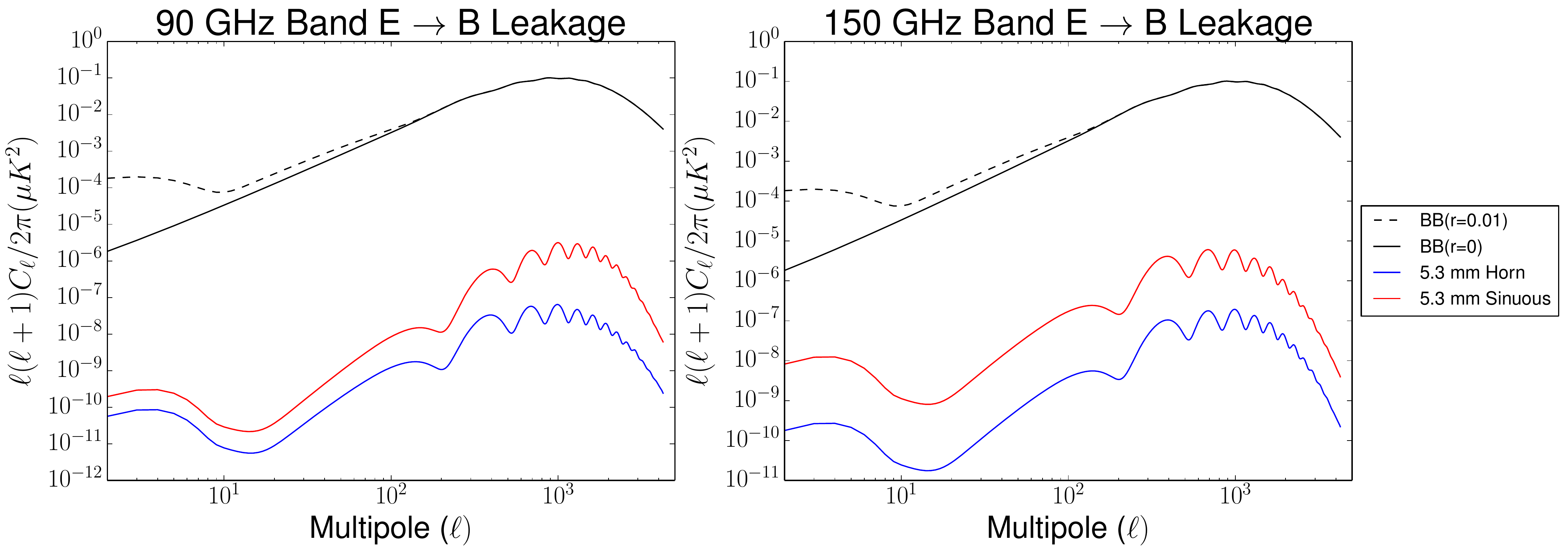}
\caption{The E-mode to B-mode leakage of a spline-profiled feedhorn design (blue) and a lenslet and sinuous antenna design (red) for a 5.3 mm pixel size on the LAT for the 90~GHz (left) and 150~GHz (right) bands. The leakage is plotted with simulated B-modes in black for $r=0$ (solid) and $r=0.1$ (dashed). The leakages of both architectures are negligible.}
\label{fig:MF_EB_leakage}
\end{figure}

For feedhorns, the main source of leakage is from ellipticity. Based on the symmetry arguments presented in Shimon, et al., 2008~\cite{Shimon_2008}, the T$\rightarrow$P leakage leakage goes primarily into E-modes, and the E$\rightarrow$B leakage is negligible. The primary source of leakage in a lenslet-coupled sinuous antenna is wobble in both the polarization and ellipticity axis. To account for the wobble in the polarization axis, the band-averaged $\delta$ beam must be rotated into the band-averaged $\epsilon$ beam until $\epsilon$ is minimized for each band, and the band-averaged $\sigma$ beam must be spatially rotated by the same angle. The resulting E$\rightarrow$B leakage is negligible, and the T$\rightarrow$P leakage goes primarily into E-modes. We note that the E$\rightarrow$B leakage after the polarization angle rotation is dominated by a monopole contribution, which could be due to imperfect rotation and/or numerical error in the simulations, so the estimated E$\rightarrow$B leakage represents an upper limit, especially at low-$\ell$. The wobble can be further mitigated through the use of a four-pixel subtraction scheme, which uses two detectors offset in polarization angle by $90^{\circ}$ for two alternating sinuous antenna rotation directions~\cite{TokiThesis}. This scheme places additional array layout constraints on the rotation direction and polarization orientation that are not present in the feedhorn and OMT architecture. Figure~\ref{fig:MF_EB_leakage} compares the E$\rightarrow$B leakage of a feedhorn (blue) and lenslet (red) for a 5.3~mm pixel pitch on the LAT using the window function method.

\begin{figure}[h!]
\centering
\includegraphics[width=0.8\textwidth]{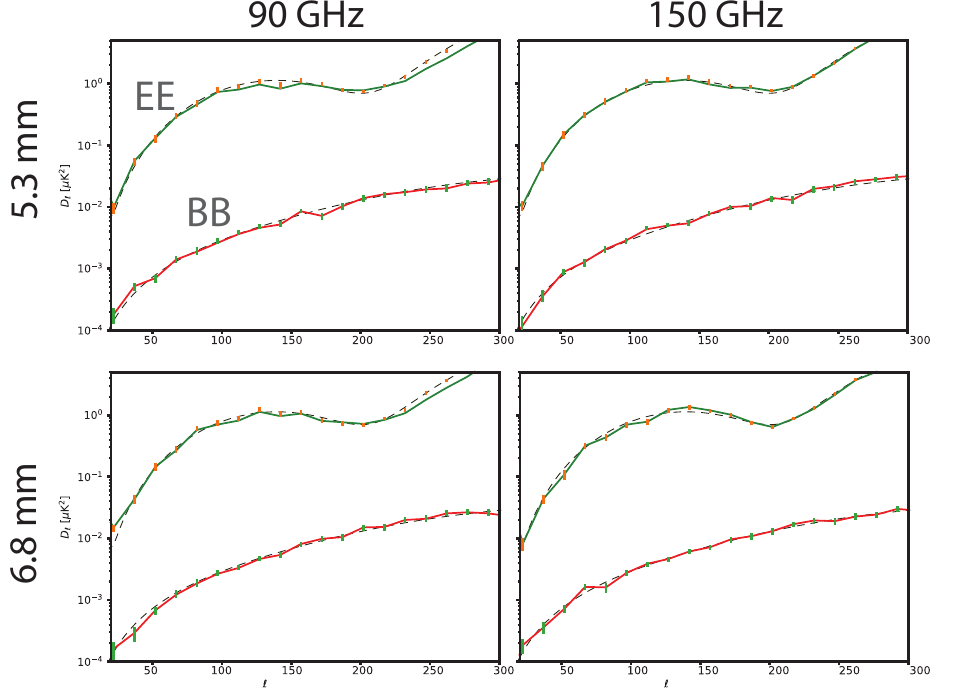}
\caption{The simulated E-mode and B-mode polarization spectra with (solid lines) and without (error bars) T$\rightarrow$P leakage from a spline-profiled feedhorn design for a 5.3~mm (top row) and 6.8~mm (bottom row) pixel size on the SAT for the 90 GHz band (left column) and 150~GHz band (right column). The dashed lines represent the E-mode and B-mode simulations with no noise variation. For these simulations, $r=0.0001$. As expected, the feedhorn leakage goes primarily into the E-mode signal and is higher in the 90~GHz band due to the waveguide cutoff of the feedhorn. The level of leakage will be further reduced by at least a factor of 10 when beam calibration is included.}
\label{fig:MF_IP_horn}
\end{figure}

\begin{figure}[h!]
\centering
\includegraphics[width=0.8\textwidth]{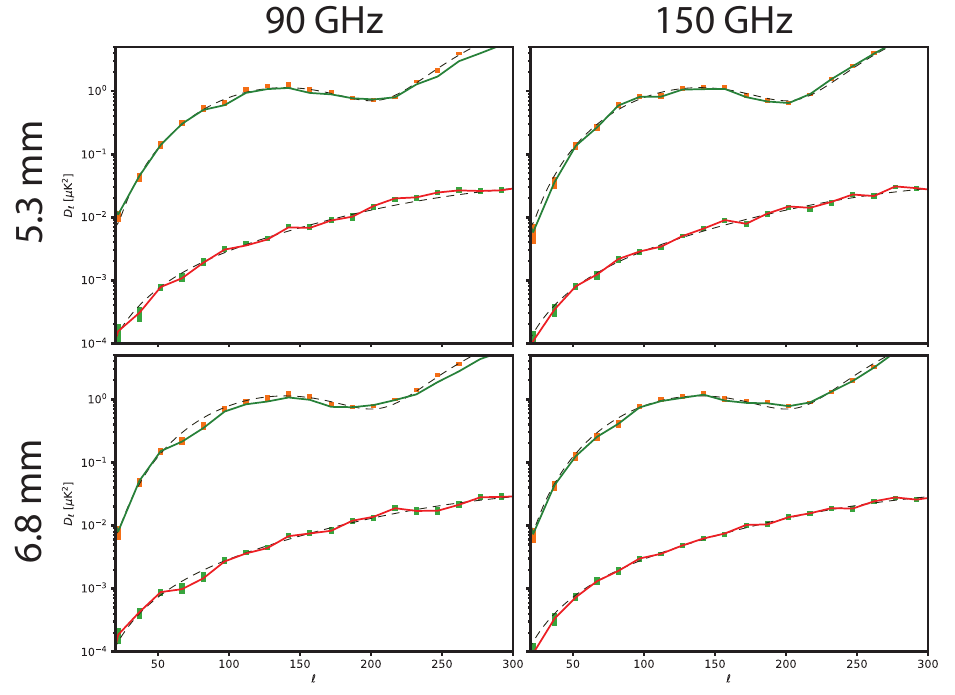}
\caption{The simulated polarization spectra with and without T$\rightarrow$P leakage from lenslet-coupled sinuous antenna designs for a 5.3~mm (top row) and 6.8~mm (bottom row) pixel size on the SAT at 90 GHz (left column) and 150~GHz (right column) with the same convention as Fig.~\ref{fig:MF_IP_horn}. The T$\rightarrow$P leakage goes primarily into E-modes.}
\label{fig:MF_IP_sinuous}
\end{figure}

The SAT has a larger stop than the LAT and subtends a larger angle as seen from the detectors. This means that more of the beam sidelobes and asymmetry make it through the stop, so the T$\rightarrow$P leakage is larger than for the LAT. To determine the relative amounts of temperature leakage that go into E-modes versus B-modes, the map-based systematics pipeline must be used. Figures~\ref{fig:MF_IP_horn} and~\ref{fig:MF_IP_sinuous} show the temperature to E-mode and B-mode leakage for the SAT for both 5.3~mm and 6.8~mm pixel size designs using the map-based simulation pipeline, and Table~\ref{tab:IP_horn_lens} shows the $\Delta\chi^2$ values between the case of no leakage and temperature to polarization leakage averaged over ten realizations for $20 \leq \ell \leq 130$. These results confirm that the T$\rightarrow$P leakage goes into mainly E-modes for both feedhorns and the lenslet-coupled sinuous antenna. We define an acceptable level of leakage as $\Delta\chi^2/10<0.5$, where the factor of ten represents the minimum suppression from beam calibration. Both the feedhorn-coupled OMT and lenslet-coupled sinuous antenna architectures have acceptable levels of T$\rightarrow$P leakage for SO both on the LAT and the SAT at both pixel sizes when beam calibration is included, and the E$\rightarrow$B leakage of both is negligible. Further, it is important to note that the SAT will employ a CRHWP, which will strongly mitigate these effects. More detailed simulations that incorporate expected levels of beam calibration, 4-pixel wobble removal, and CRHWP demodulation for SO are currently underway.  Both the feedhorn-coupled OMT antenna and lenslet-coupled sinuous antenna meet the requirements for the LAT, and, while the requirements are more stringent for the SAT, both technologies will exceed the requirements once the effects of the CRHWP and beam calibration are included.

\begin{table}[h!]
\centering
\label{tab:IP_horn_lens}
\begin{tabular}{|c|c|l|l|l|l|}
\hline
\multirow{2}{*}{Architecture} & \multirow{2}{*}{Pixel Size (mm)} & \multicolumn{2}{c|}{$\Delta\chi_{EE}^2$}                        & \multicolumn{2}{c|}{$\Delta\chi_{BB}^2$}                        \\ \cline{3-6} 
                              &                                  & \multicolumn{1}{c|}{90 GHz Band} & \multicolumn{1}{c|}{150 GHz Band} & \multicolumn{1}{c|}{90 GHz Band} & \multicolumn{1}{c|}{150 GHz Band} \\ \hline
Feedhorn+OMT                          & 5.3                              & 1.84           & 0.79           & 1.02           & 0.41           \\ \hline
Feedhorn+OMT                          & 6.8                              & 1.20           & 0.12           & 0.41          & 0.21          \\ \hline
Lenslet+Sinuous                       & 5.3                              & 2.11          & 0.78           & 1.17           & 0.39            \\ \hline
Lenslet+Sinuous                       & 6.8                              & 1.85          & 0.86           & 0.82           & 0.26             \\ \hline
\end{tabular}
\caption{The $\Delta\chi^2$ values between the Gaussian beam case with no systematic effects and the simulated beams for feedhorns or the lenslet and sinuous antennae cases averaged over ten realizations on the SAT with 5.3~mm and 6.8~mm pixel sizes. We note that beam calibration can reduce the $\Delta\chi^2$ values by at least a factor of ten. This table only includes T$\rightarrow$P leakage and does not include E$\rightarrow$B leakage, which is negligible for all cases.}
\end{table}

\section{Non-linearity}\label{sec:nonlinearity}

Thus far, we have discussed generic polarization-sensitive detectors with millimeter-wave couplings via feedhorns or lenslets. In this section, we discuss detector nonidealities specific to the case of transition-edge sensor (TES) bolometers. These devices have become standard in the field of CMB instrumentation~\cite{Henderson_2016,Inoue_2016,Benson_2014,TEH_2014} as they have achieved NETs dominated by photon noise, the fundamental limit on per-detector sensitivity. Detectors that achieve this performance are called ``background-limited.'' Detailed models of the linear response of the TES bolometer to small signals are well-studied\cite{IrwinHilton,Maasilta:2012bx}, and have been broadly confirmed by detailed characterization studies~\cite{George:AlMnTES,Hubmayr:SPIDER}.

The above models are explicitly derived assuming small deviations around stable operating points for the TESes themselves.  However, it is also known that the conditions describing the operating point of a device (e.g. the TES resistance, the bias power sourced by the TES resistance, the sensitivity to changes in temperature) slowly vary during observations. For ground-based observatories, this is often attributed to the change in incident radiated power on the bolometers due to the changing atmosphere or other environmental factors. Though these variations are seen as additional long-timescale variance with $1/f^{a}$ spectra in the timestreams of the bolometers, we also expect that these changes are sufficiently large to alter the TES bolometer's response to the sky signal.

Since we usually only have access to the apparent size of atmospheric fluctuations on long timescales through the timestreams of the TES bolometers themselves, we are not able to extract unbiased measures of the atmospheric power and measure this non-linearity directly. However, we consider the inherent non-linearities of TES bolometers, drawing from the observations of T$\rightarrow$P leakage induced by this effect in experiments with CRHWPs~\cite{Takakura:2017ddx,Didier:EBEX_HWP}. Discussion of non-linearity-induced leakage for the SATs are described in another SO systematics study in these proceedings~\cite{salatino18}. Since the non-linearity effect is completely general, in this paper, we consider the case of such T$\rightarrow$P leakage on experiments that recover polarization information through the pair-differencing of bolometer pairs with orthogonal polarization sensitivity. While some CMB experiments without CRHWPs do not explicitly pair-difference, we consider the pair-differencing case for the SO LAT in our simulations. Specifically, we focus on the 150~GHz band of the SO multichroic 90/150~GHz detector array.

First, we assume the following equation that transforms our input signal timestream $d(t)$ into the distorted $d_{\mbox{\scriptsize NL}}$~\cite{Takakura:2017ddx}:
\begin{equation}
d_{\mbox{\scriptsize NL}} = \left[ 1 + g_1 d(t) \right] d( t - \tau_1 d(t) ),
\label{eq:nl_observe}
\end{equation}
where for completely linear, idealized detectors, $g_1$ and $\tau_1$ would be zero. This is a small-signal approximation to a first-order nonlinear gain affected by signal level, where the approximation is in the effect of $\tau_1$ on a timestream. Calculations of the size of the non-linearity parameters $g_1$ and $\tau_1$ for the case of TES bolometers and CRHWP experiments \cite{Takakura:2017ddx} include the polarization modulation frequency as a parameter. We instead take this frequency to zero, which causes negligible change in the estimated size of the non-linearity terms. We use estimated SO bolometer parameters to determine the approximate sizes of $g_1$ and $\tau_1$. Specifically, we focus on detector parameters determined by sensitivity studies for the 150~GHz channel on the LAT. Relevant parameters include target saturation power $P_{\mbox{\scriptsize sat}}$ = 6.3~pW, target TES normal resistance $R_{\mbox{\scriptsize N}}$ = 8~m$\Omega$, and target thermal time constant $\tau = C/G$ = 8~ms, where $G$ is the conductance between the TES bolometer and the thermal bath, and $C$ is the heat capacity of the bolometer. We assume that the loop gain $\mathscr{L}$ (introduced below) is approximately 15. The resulting characteristic values of the non-linearity parameters assumed for these studies are $g_1 =$ -0.44~\%/K and $\tau_1 =$ 0.005~ms/K. Again, these numbers are specific to the 150~GHz LAT configuration. These values will shift with design changes, particularly in adjustments to $P_{\mbox{\scriptsize sat}}$ and the target time constant from changes to $G$.

Importantly, both terms are inversely proportional to powers of the Joule power dissipated in the TES,  $P_{\mbox{\scriptsize bias}}$, and the TES loop gain $\mathscr{L}$:
\begin{equation}
\mathscr{L} = \frac{P_{\mbox{\scriptsize bias}} \alpha}{G T_c},
\label{eq:loop_gain_def}
\end{equation}
where $\alpha$ parametrizes the TES sensitivity to temperature fluctuations, $G$ is the bolometer thermal conductance to bath, and $T_c$ is the TES critical temperature, assumed to be the bolometer temperature during operation. Increasing bias power reduces $g_1$ as $P^{-2}_{\mbox{\scriptsize bias}}$ and $\tau_1$ as $P^{-1}_{\mbox{\scriptsize bias}}$.~\cite{Takakura:2017ddx}

For our simulations, we require realistic atmospheric signals to be observed by nonlinear detectors, where the difference in the induced gain variations of the two bolometers in a pair will cause T$\rightarrow$P leakage. We simulate correlated noise based on a model of $N_\ell$, the noise angular power spectrum, determined from experimental data acquired by similar, existing telescopes observing from the same high-altitude site in Chile. The studied $N_\ell$ curves do not represent our estimates of the SO noise performance. 

For this initial model, we describe correlated noise as:
\begin{equation}
n_{\mbox{\scriptsize corr}} (f) = A \left( \frac{f_p}{f} \right)^{a},
\label{eq:corr_noise}
\end{equation}
where $A$, representing the power at $\ell =$ 1000 in the $N_\ell$ curve, and $a$, the power-law exponent of the noise angular power spectrum, are drawn randomly from a normal distribution for each simulated CES, and the per-CES estimation of $f_p$ is discussed below, but the $\ell$ pivot scale is fixed at $\ell =$ 1000. This equation does not specify the source of the noise in a given observation, but the primary source of long-timescale variance in the autocorrelation and cross-correlation of detectors is the atmosphere.

The $N_\ell$ model discussed above defines the mean values of $A$ and $a$, though we apply an empirical reduction of $A$ by a factor of 10. We believe that the need for this reduction may point to a conversion factor from $\ell$ to frequency, which we plan to estimate in future simulations. The central values used in the simulations are $A =$ 350 and $a =$ 3.5. The pivot frequency $f_p$ is estimated from the pivot value of the $N_\ell$ curve according to a linear relation between sky scale and timestream frequency $f$:
\begin{equation}
\ell \sim 360^\circ \times \frac{f}{v_{\mbox{\scriptsize scan}} \cos (\theta)},
\label{eq:ell2freq}
\end{equation}
where $v_{\mbox{\scriptsize scan}}$ is the telescope azimuth scan speed in $^{\circ}$/s and $\theta$ is the CES elevation. We emphasize that this is an empirical approximation.

As mentioned in Sec.~\ref{subsec:s4cmb}, $t_{\mbox{\scriptsize corr}}$ and $n_{\mbox{\scriptsize cloud}}$ are used to define additional aspects of the correlated noise. The parameter $t_{\mbox{\scriptsize corr}}$ describes the time scale over which the correlated component of the noise is generated. After each interval of this length, a new draw of the correlated noise is performed, which implements a coherence timescale for the correlated noise. The parameter $n_{\mbox{\scriptsize cloud}}$ defines a number of localized subarrays within the array over which the atmosphere has common phases of the Fourier components, whose power spectrum obeys Eq. \ref{eq:corr_noise}. We set these to $t_{\mbox{\scriptsize corr}} = 5$~ minutes and $n_{\mbox{\scriptsize cloud}} = 3$ for our small-scale simulations, which are the nominal defaults in \textit{s4cmb}. We do not yet have a model for estimating how this parameter might itself vary with atmosphere precipitable water vapor (PWV), or for the expectation that $t_{\mbox{\scriptsize corr}}$ is driven by environmental factors like wind speed.

Using estimates of the dependence of the LAT array sensitivity on PWV from the BoloCalc sensitivity calculator~\cite{hill18}, we vary the array (and per-detector) white noise level as a function of a random PWV. We scale from the estimated sensitivity of the entire set of 150~GHz-channel bolometers on the LAT, and further reduce this number assuming a total of six months observing at 20\% efficiency.

In defining a PWV value, we draw from an approximation to the observed distribution of PWV data measured during POLARBEAR-1 observations. In addition, we account for the sensitivity calculator's estimates on how varying PWV changes the bias power on the TES. This involves scaling the estimated values for $g_1$ and $\tau_1$ based on the equations in Takakura, et al., 2017~\cite{Takakura:2017ddx} by the appropriate factor of $P_{\mbox{\scriptsize bias}}$ to define mean non-linearity parameters for each CES. Individual bolometer non-linearity parameters are then drawn from a normal distribution with a standard deviation of 10\% of the per-CES mean value.

These parameters are then used to distort each detector's timestream, which consists of CMB signal, white noise, and correlated noise. We find that the lag induced by $\tau_1$ is often less than half a sample time given the simulation sampling rate of 32~Hz. Since we round the induced time lag to an integer index shift, this means $\tau_1$ is not a factor in this simulation.

However, a more physically relevant timescale for $\tau_1$ is the beam crossing time, which for the 1.4' beam size of the 150~GHz channel on LAT, the 2.5$^{\circ}$/s scan speed, and a conservative elevation of 50$^{\circ}$, is equal to 15~ms, or 70~Hz. For $\tau_1$ to reach 1\% of this timescale requires excursions of 30~K in the time domain. Such a baseline change is extreme but possible for the correlated noise powers discussed here. Proper treatment of the effects of $\tau_1$ will involve simulating fast-sampled timestreams nearer the SO LAT sampling rate of 400~Hz and incorporating nonlinear observations, which will be done for future full-scale simulations.

When simulating SO LAT observations, we study both the ``shallow" and ``deep" observing strategies laid out as \textit{s4cmb} defaults. The former runs large-throw scans for wider sky coverage, relevant for maps to be used in cross-correlation studies with other tracers of large-scale structure and measurements of the lensing potential. The latter observes a smaller area with deeper coverage. 

\begin{figure}[t]
\centering
\includegraphics[width=0.8\textwidth]{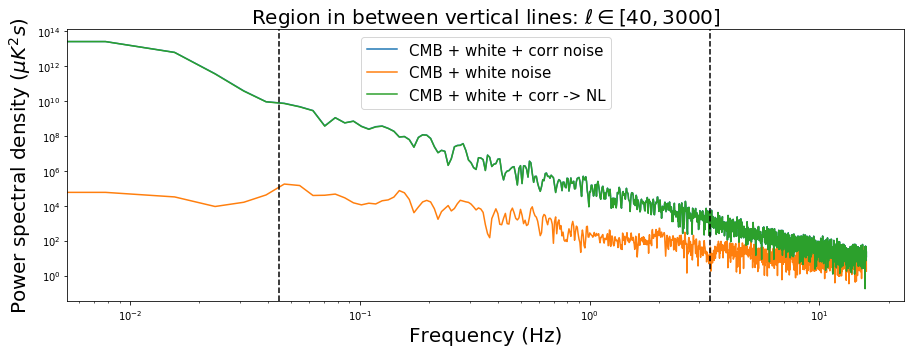}
\caption{Simulated noise spectral densities for one detector for a length of time $t_{\mbox{\scriptsize corr}}$ at the start of a CES. The three solid lines correspond to three observing schemes: CMB+white noise (orange), CMB+white noise+correlated noise (blue, not visible beneath green), and the nonlinear observed data of the CMB+white noise+correlated data (green). The above shows the characteristic size of the atmospheric fluctuations assumed and their approximate knee at $\sim$ few Hz. The vertical dashed lines indicate the approximate $\ell$ range quoted in the plot title as it maps on to noise frequency with a 2.5$^{\circ}$/s scan speed for the observation.}
\label{fig:corr_noise_ex}
\end{figure}

Figure~\ref{fig:corr_noise_ex} shows the noise spectral density in CMB temperature units for one bolometer within one CES in the deep scanning strategy. The three cases shown are: the CMB with white noise, adding correlated noise, and distorting the timestream with the non-linearity function. The effect of the non-linearity is not visible in the plot, but it is apparent in the reconstructed maps.

\begin{figure}[t!]
\centering
\includegraphics[width=0.7\textwidth]{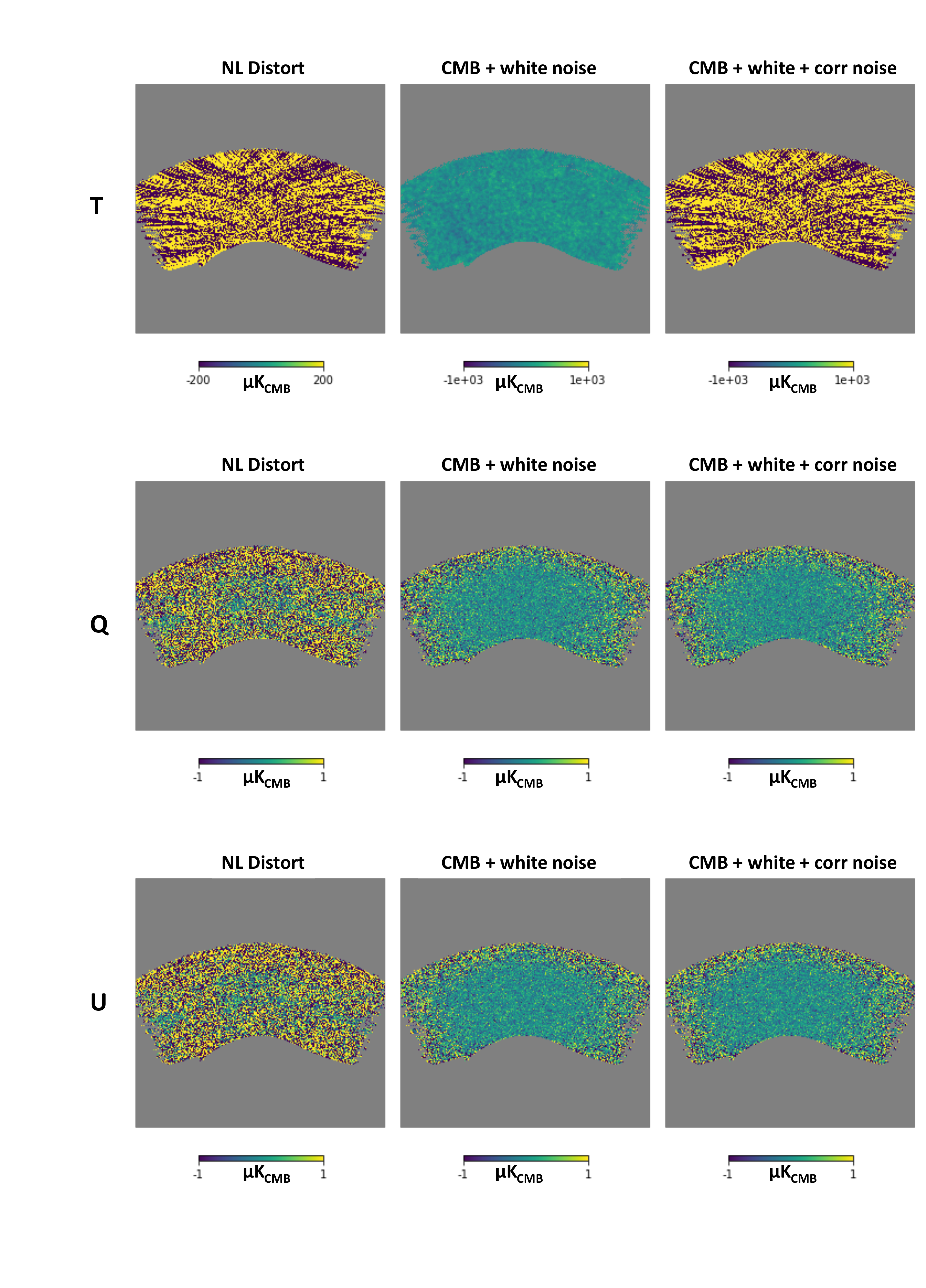}
\caption{Maps of Stokes parameters $I,Q,U$ (in rows) and the simulation types used to generate them (in columns) for the deep observing scheme implemented in \textit{s4cmb}. These simulations used a small number of bolometers on a square grid focal plane, but the simulation pipeline can be run for the full SO focal plane as the instrument design progresses. The middle column shows the simulated CMB maps with white noise only. The maps in the right column add correlated noise, and the maps in the left column take the CMB+white noise+correlated noise maps and add non-linearity.  Thus, we see we have induced a $\sim$1~$\mu$K polarization signal into our maps via non-linearity. Correlated noise alone does not leak T$\rightarrow$P, as the residuals of subtracting the simulated white noise map from the correlated noise map are negligible in $Q$ and $U$. The area of the observed patch is $\sim$1\% of the sky.}
\label{fig:nl_leak_map_deep}
\end{figure}

We produce our output sky maps of the Stokes $I$, $Q$, and $U$ parameters separately for the three simulated cases. These are shown in Fig.~\ref{fig:nl_leak_map_deep} for the deep observing strategy. In this figure, 32 SO-like detectors are simulated in observations while the total array sensitivity is held constant, according to Eq. \ref{eq:perdet_sensitivity} in Sec. \ref{sec:analysis_frameworks}. The noise reduction of a six-month observing campaign is incorporated by scaling down this sensitivity. In this case, we have intentionally set CMB $Q$ and $U$ components to be zero so that any $Q$ and $U$ observed in the maps can be considered leakage. The non-linearity-induced gain mismatches cannot be neglected. The evident amplitude of these signals in Fig.~\ref{fig:nl_leak_map_deep} indicates that further study is necessary. In future work, we will determine how this leakage might average down across larger detector counts as well as exploring the effects of the leakage on the angular power spectrum. The coupling of gain fluctuations driven by the atmospheric intensity signal on long timescales through the TES bolometer's inherent non-linearity can be pernicious. We plan to establish an acceptable upper bound on the leakage by studying full SO-scale simulation results in spectral space with $P_{\mbox{\scriptsize bias}}$ allowed to vary to scale $g_1$ and $\tau_1$.

\section{Gain Drifts}\label{sec:gain_drift}
In this section, we present a second source of gain drifts: long-timescale instability of the thermal bath temperature during observations of a scanning telescope. We present the relevant equations governing this sky-independent leakage, and discuss the possible residuals induced by such an effect in both the SO SAT and LAT cameras.

The expression in the frequency domain for the power-to-current responsivity $s_I$ of a TES bolometer in the limit of stiff voltage bias, negligible sensitivity to current fluctuations, and negligible inductance in series with the TES is given by:~\cite{IrwinHilton}
\begin{equation}
 s_I \sim - \frac{1}{V_{\mbox{\scriptsize TES}}} \frac{\mathscr{L}}{\mathscr{L}+1} \frac{1}{1 + i \omega \tau_{\mbox{\scriptsize eff}}}\,\, ,
\label{eq:resp_tes_def}
\end{equation}
where $V_{\mbox{\scriptsize TES}}$ is the bias voltage supplied by the readout electronics, $\omega$ is the angular frequency ($\omega = 2 \pi f$), and $\tau_{\mbox{\scriptsize eff}}$ is the effective detector time constant. Given the conditions above, we approximate $\tau_{\mbox{\scriptsize eff}}$ as:
\begin{equation}
\tau_{\mbox{\scriptsize eff}} \sim \frac{\tau}{1 +\mathscr{L}}.
 \label{eq:tau_eff_def}
 \end{equation}
Again, $\tau$ is the pure thermal time constant of the TES, $\tau = C/G$.

Implicit in Eq.~\ref{eq:resp_tes_def} are dependencies on the bath temperature, \tbath, of the bolometer, which we usually assume to be the measured temperature of the mechanical elements of the TES bolometer array. To determine the dependence on \tbath, we first consider the saturation power, $P_{\mbox{\scriptsize sat}}$, of the TES bolometer. If the sum of all powers incident on the TES exceeds this value, the TES is driven normal and will not be sensitive to changes in optical power. However, during operation when the TES is on its resistive transition and thus close to the normal state, the total power on the TES can be approximated as the saturation power. We express $P_{\mbox{\scriptsize sat}}$ as a difference of powers of the TES temperature $T_c$ and the bath temperature, with conversion factor $\kappa$:
\begin{equation}
P_{\mbox{\scriptsize sat}} = \kappa \left( T^n_c - T^n_{\mbox{\scriptsize bath}} \right).
\label{eq:psat_def}
\end{equation}

Since \tbath is generally assumed constant, we often take $P_{\mbox{\scriptsize sat}}$ to be constant. However, we may introduce the quantity $\xi$:
\begin{equation}
\xi = \frac{dP_{\mbox{\scriptsize sat}}}{dT_{\mbox{\scriptsize bath}}} = -n \kappa T^{n-1}_{\mbox{\scriptsize bath}},
\label{g_def}
\end{equation}
which is analogous to the thermal conductance to bath $G$.

We express $\mathscr{L}$ in terms of $P_{\mbox{\scriptsize sat}}$ using $P_{\mbox{\scriptsize bias}} + P_{\gamma} = P_{\mbox{\scriptsize sat}}$, where $P_{\gamma}$ is the incident optical power and is independent of \tbath. The derivative of $P_{\mbox{\scriptsize bias}}$ (or $P_{\mbox{\scriptsize sat}} - P_{\gamma}$ ) with respect to \tbath is then given by $\xi$. The derivative of the loop gain with respect to \tbath is then:
\begin{equation}
\frac{d \mathscr{L}}{d T_{\mbox{\scriptsize bath}}} = \mathscr{L} \frac{\xi}{P_{\mbox{\scriptsize bias}}}.
\label{eq:loop_deriv}
\end{equation}

Finally, we take the derivative of $s_I$ with respect to \tbath. We do so by only looking at the chain-rule derivatives depending on $\mathscr{L}$ and assuming $V_{\mbox{\scriptsize TES}} \sim$ constant. The recovered signal, $\delta P'_{\gamma}$, for the input signal $\delta P_{\gamma}$ undergoing a given bath temperature fluctuation signal $\delta T_{\mbox{\scriptsize bath}} (t)$ is then:
\begin{equation}
\delta P'_{\gamma} = \left( 1 + \frac{d s_I / d T_{\mbox{\scriptsize bath}} }{s_I} \; \delta T_{\mbox{\scriptsize bath}} \right) \delta P_{\gamma} = ( 1 + \rho\;\delta T_{\mbox{\scriptsize bath}} )\delta P_{\gamma}.
\label{eq:recovered_power_fl}
\end{equation}

The gain fluctuation fraction $\rho$ can be written simply as:
\begin{equation}
\rho = \xi \left( \frac{s_I}{I_{\mbox{\scriptsize TES}}} - \frac{1}{P_{\mbox{\scriptsize bias}}} \right).
\label{eq:rho_def}
\end{equation}
The two terms are exactly equal for $s_I = - 1 / V_{\mbox{\scriptsize TES}}$, which is the case for $\mathscr{L} \to \infty$. Thus their difference scales roughly as $\mathscr{L} / ( \mathscr{L} + 1)$.

\begin{figure}[h]
\centering
\includegraphics[width=0.92\textwidth]{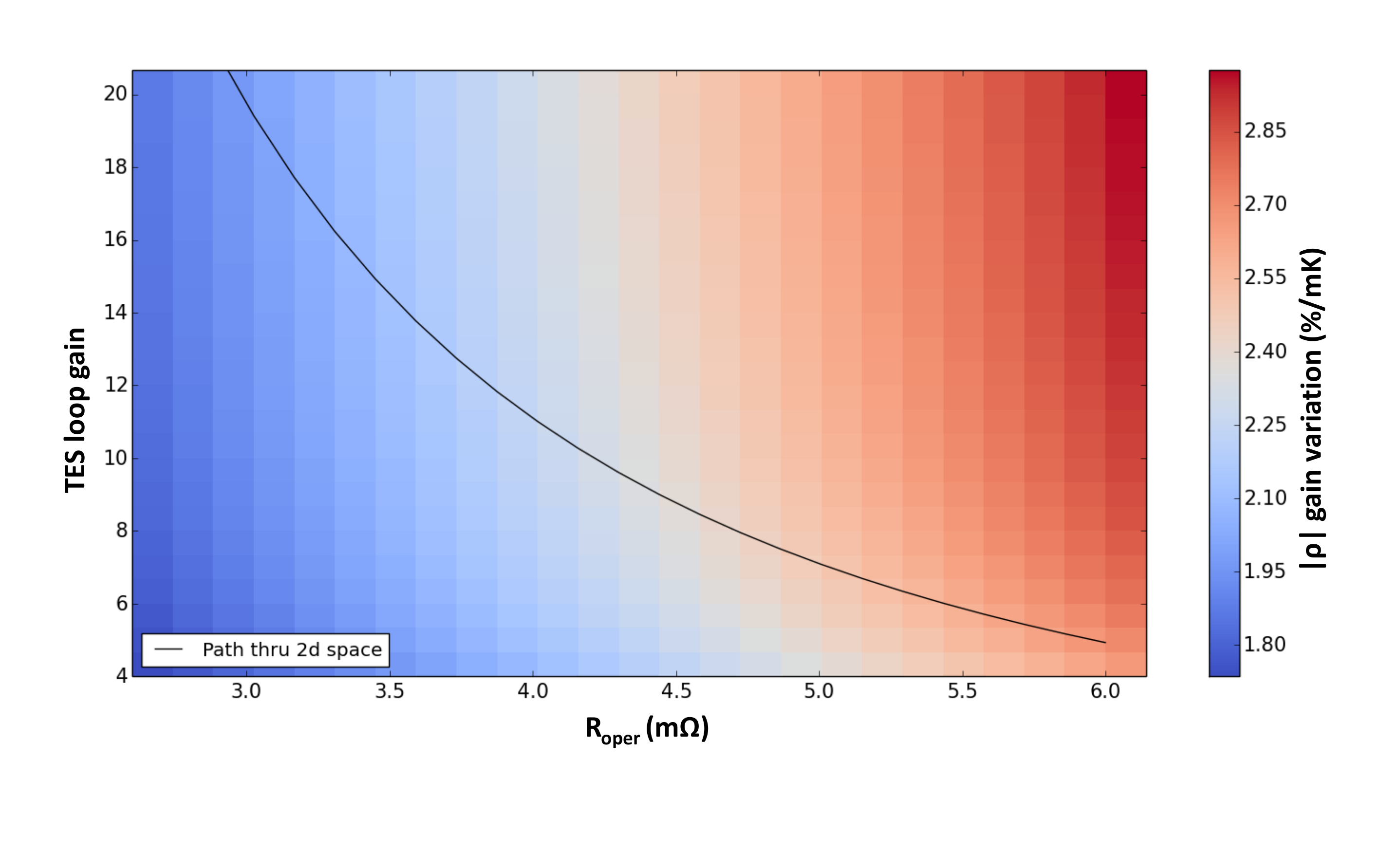}
\caption{Leakage coefficient $\rho$ as a function of loop gain $\mathscr{L}$ and TES resistance $R$. As $R$ changes, we hold $V_{\mbox{\scriptsize TES}}$ and allow $P_{\mbox{\scriptsize bias}}$ to vary. The parametric curve indicates the path an SO-like TES under constant voltage bias would take through the space, letting $P_{\mbox{\scriptsize bias}}$ vary as described above and assuming a simple inverse dependence of the TES sensitivity $\alpha$ on $R$.}
\label{fig:tbath_leak_matrix}
\end{figure}

In Fig.~\ref{fig:tbath_leak_matrix}, we show the absolute value of $\rho$ in units of \%/mK assuming SO-like TESes and focal plane properties for a fluctuation frequency of 1~Hz. Here, ``SO-like TES" means that we have used the same sets of bolometer-defining numbers used for the simulations of Sec. \ref{sec:nonlinearity}, as they apply to the 150~GHz-band detectors on the LAT. As for selecting 1~Hz, for frequencies below the TES $f_{\mbox{\scriptsize 3dB,eff}}$ defined by $f_{\mbox{\scriptsize 3dB,eff}} = 1/2 \pi \tau_{\mbox{\scriptsize eff}}$, the quantity $|\rho|$ depends very weakly on frequency. We find that the gain variations can be as significant as a few \%/mK.

The frequency dependence of $\rho$ comes entirely from the term dependent on $s_I$. In this model, we assume that the leakage is flat even to low frequencies. However, we fully expect that second-order effects beyond changing loop gains and bias powers may be relevant at these frequencies. In addition, the complex interplay of low-frequency, high-variance modes in $P_{\gamma}$ due to atmospheric fluctuations and these bath temperature fluctuations in the context of maintaining total power $P_{\mbox{\scriptsize sat}}$ on the TES has explicitly not been considered.

In the case of pair-differenced observations on an instrument similar to the LAT, the relevant parameter for leakage is $\delta \rho$, the difference between the $\rho$ values of detectors in a pair. Though we may imagine bath temperature fluctuations to be pernicious as a common mode, this effect will be reduced assuming even modest array uniformity in electrical bias values and the thermal parameters informing $\xi$.

Minimizing $\rho$ minimizes this systematic effect. Making $\rho$ small requires low responsivity, high TES current, and/or large $P_{\mbox{\scriptsize bias}}$. The latter two are consistent, and also ensure a more linear detector through the loop gain $\mathscr{L} \propto P_{\mbox{\scriptsize bias}}$. High TES current also means higher TES voltage for a given operating resistance, and thus smaller responsivity. However, increasing these powers to minimize the effect has an upper limit based on $P_{\mbox{\scriptsize sat}}$, which is usually optimized to achieve the best TES bolometer sensitivity. Additionally, decreasing responsivity affects the total contribution of current noise sources in the TES bolometer to the total noise quadrature sum, directly lowering signal-to-noise. Therefore, we plan to carefully study this leakage by: (1) generating expected spectra of the thermal fluctuations given the telescope mount design, scan architecture, and cryogenics; (2) feeding this in to a time-domain simulation scheme that can include these gain variations, and (3) projecting this analysis forward to power spectra space. Points (2) and (3) can be achieved by using \textit{s4cmb} given realistic assumptions about bath temperature variations as well as TES thermal and electrical parameter variations within an array. Work on these simulations is ongoing, and will help inform the detector design optimization.
\section{Crosstalk}\label{sec:crosstalk}

We next consider the impact of detector crosstalk induced by our detector readout architecture. Here crosstalk refers to any spurious coupling of one detector's signal into the measured signal of another detector. To study this, we generate a model of crosstalk based on our knowledge of the electronics that are used to readout our TES detector array. We write down a model of the crosstalk mechanism that is applicable to both our fiducial readout architecture, the microwave multiplexer ($\mu$mux)~\cite{Mates_Thesis_2011}, and our backup readout architecture, the Digital Frequency Domain Multiplexer (DfMUX)~\cite{Dfmux_readout_2012}. 

\subsection{Crosstalk Model from Readout Architecture} \label{CrosstalkModel}

Both of these systems are based on frequency division multiplexing, in which each TES signal is read out at a unique frequency tone. In both approaches, we digitally synthesize a frequency comb of many tones at room temperature. We then interrogate a set of resonators that match the synthesized comb frequencies. As our bolometers' resistances change with changes in optical power, the resonances are either frequency modulated ($\mu$mux ~\cite{Mates_Thesis_2011}) or amplitude modulated (DfMUX ~\cite{Dfmux_readout_2012}). This modulated comb is then returned to a set of room temperature electronics for digital demodulation to extract the bolometers' time-ordered data (TOD). This allows us to readout many detectors (equal to the number of tones in the comb $\sim$2000 for $\mu$mux and $\sim$70 for DfMUX) with $\sim$1 pair of conductors to minimize the conductive loading and reduce the complexity of interconnects from room temperature to the sub Kelvin bolometer stage. This will enable SO to field a significantly larger number of detectors ($\sim 30,000$ for both the small and large aperture cameras) than current experiments like Advanced ACTPol and POLARBEAR. 

These architectures are necessary to decrease thermal loading on the cryogenic components of our instruments; however, they will inevitably induce a non-zero level of crosstalk between detectors. The crosstalk mechanism common to both readout architectures comes from the fact that the frequency response of each resonator has a Lorentzian line shape. This means that there is a non-zero response to a given frequency $f_n$ at a neighboring frequency $f_{n+k}$ where $n$, and $k$ are integers in the range $1 < n, n+k < N$ with $N$ being the total number of tones used in the readout. This means that as the tone at $f_n$ is amplitude or frequency modulated by a bolometer signal, the neighboring detectors $f_{n+k}$ will also be modulated producing a crosstalked signal. 

The details of this calculation for DfMUX are given in Dobbs, et al., 2012~\cite{Dfmux_readout_2012} and the resulting crosstalk leakage coefficient can be written as following:
\begin{equation}
\frac{\Delta I_{n\pm k}}{\Delta I_n}\simeq \frac{-R_{\rm{bolo}}^2}{(2|\omega_{n+k}-\omega_n|L)^2}\,\, ,
\label{DfMUXXtalk}
\end{equation}
where $I_n$ is the current through bolometer $n$ at frequency $f_n$, $R_{bolo}$ is the bolometer resistance, $L$ is the resonator inductance, and $|\omega_{n+k}-\omega_n|$ is the separation in frequency between channel $n+k$ and channel $n$.

For $\mu$Mux, the bolometer TOD is encoded in changes in the imaginary part of the microwave transmitted power at a given resonance frequency, $\rm{Im}[S_{21}]$. Following Mates, 2011~\cite{Mates_Crosstalk, Mates_Thesis_2011}, the crosstalk can be expressed in a form similar to Eq.~\ref{DfMUXXtalk}
\begin{equation}
\frac{\rm{Im}[\Delta S_{21,n+k}]}{\rm{Im}[\Delta S_{21,n}]} = \frac{\omega_n\rm{BW}}{32|\omega_{n+k}-\omega_n|^2 Q_{c,n}}\,\, ,
\label{umuxXtalk}
\end{equation}
where $\rm{Im}[\Delta S_{21}]$ is the change in the imaginary part of the transmission through the microwave multiplexer, $BW$ is the resonator bandwidth, and $Q_c$ is the coupling quality factor. 

To quantify the systematic contamination in observed maps, we assume that this model is both linear and time-independent. The most general way to write such a model is as a mixing matrix $L_{ij}$ that maps true optical power in detector $i$, $d^i$ to observed signal in detector $j$, $\tilde{d}^j$ following the equation:

\begin{equation}
\tilde{d}^j = d^j + \sum_{i \neq j} L_{ij} d^i\,\, .
\end{equation}

We can recast the equations derived from the circuit models into crosstalk matrix components $L_{ij}$. From these two equations, we see that both of these crosstalk mechanisms scale as $1/\Delta f^2$, where $\Delta f = f_{n+k}-f_n$. This is the first scaling that we have implemented into \textit{s4cmb}. We refer to this crosstalk term as $L^{\rm{in}}_{ij}$, since it is the crosstalk within a comb of readout tones. By design the magnitude of this term is expected to be as low as 0.03\% for $\mu$mux and 0.3\% for DfMUX and from measurement this may pessimistically be as high as 0.3\% for $\mu$mux and 3.0\% for DfMUX. In Sec.~\ref{Implementation}, we discuss implementing $L^{\rm{in}}_{ij}$ into \textit{s4cmb}. There we lump all of the constants in Eqs.~\ref{DfMUXXtalk} and~\ref{umuxXtalk} into a single parameter $\alpha$ that sets the nearest frequency neighbor crosstalk amplitude between 0.03-3.0\% and then scales the crosstalk amplitude to more distant neighbors by 1/$\Delta f^2$.

We have also implemented a constant level of crosstalk between all detectors on a given readout line (a superconducting quantum interference device (SQUID) with $\sim$70 detectors for DfMUX or a low noise amplifier with $\sim$2000 detectors for $\mu$mux). This crosstalk can be introduced from inductive crosstalk in the SQUID input coils or wiring between the cryogenic multiplexer and the room temperature demodulator. This mechanism simulates the effect of crosstalk between pixels separated by large angular scales on the sky. We expect this crosstalk level to be very small based on lab measurements and have implemented an upper bound of $L_{ij} = 0.01\%$. We refer to this crosstalk term as $L^{\rm{out}}$ since it involves crosstalk to detectors that are not necessarily adjacent in frequency.

It should be noted that the DfMUX system has a secondary crosstalk mechanism that is due to a stray inductance in the system and scales as $f_n/\Delta f$. With linearly spaced readout frequencies in the comb, this crosstalk term increases as a function of frequency. To make this effect sub-dominant, the comb is generated with logarithmic frequency spacing. Since the primary SO architecture is $\mu$mux, we have chosen not to simulate this effect.

Recent results from Dober, et al., 2017~\cite{Dober2017} show that the measured crosstalk on the current SO $\mu$mux chip design is $\sim$10 times larger than the expected level from the Lorentzian tail leakage alone. This could be explained by a nearest neighbor inductive crosstalk on the multiplexer chips. This is currently being investigated and may be mitigated with improved chip layouts. Once this mechanism, its underlying source, and its parametric dependence are understood, it will be incorporated into our model.

Lastly, we note that the simulation framework itself is agnostic to the physical crosstalk mechanisms, so a crosstalk mixing matrix $L_{ij}$ measured in a lab can be directly simulated. As the design within SO matures, future simulations of crosstalk will be based on direct measurements of the mixing matrix. If the crosstalk matrix $L_{ij}$ is measured and stable, the 1st-order effects of crosstalk can be perfectly removed, and we are left only with secondary effects due to imperfect removal due to calibration error and/or time-variability. We plan to extend this framework to assess the spurious signals generated by these secondary effects in future work.

\subsection{Implementing a Crosstalk Model into Time-Domain Simulations} \label{Implementation}

Our simulations are based on a toy model of a focal plane containing 6,272 detectors. In this model, the readout frequency increases left to right and top to bottom as in written text. In this model, adjacent bias frequencies are adjacent physically. This layout may significantly overstate the crosstalk level because the contamination will not average down across detectors, so we treat it as an upper limit and note that the detector count and focal plane geometry of the SO LAT and SAT instruments will be different. A more realistic hardware layout will be included in future studies as the SO design matures, and these improved crosstalk simulations will be used to determine the optimal layout for minimizing crosstalk.

The crosstalk matrix $L_{ij}$ is constructed as follows in the time-domain simulations. For the Lorentzian tail component, $L_{ij}^{\rm{in}}$, each bolometer timestream $d^i$, has a nearest frequency neighbor crosstalk leakage coefficient $\alpha_i$ drawn from a normal distribution with mean, $\mu$, and standard deviation, $\sigma$. We use Gaussian distribution since we do not yet have a better model for variability. We simulate both a nominal case with $\mu$ = -0.3\% and $\sigma$ = 0.1\% and a pessimistic case with $\mu$ = -3.0\% and $\sigma$ = 1.0\%. These values are realistic for both $\mu$mux and DfMUX as discussed in section \ref{CrosstalkModel}. It should be noted that $\mu$mux has demonstrated crosstalk levels lower than our nominal case with $\mu$ $\sim$ -0.03\%\cite{Dober2017}. The crosstalk matrix $L_{ij}$ is assumed to be time-independent.

Bolometers  are grouped based on the multiplexing factor $N_{\rm{mux}} \sim 2000$ detector units for $\mu$Mux or $N_{\rm{mux}} \sim 68$ detector units for DfMUX. This is used to define the set of pairs of bolometers with non-zero crosstalk. The matrix elements of $L_{ij}$ for a bolometer $b_i$ talking with another bolometer $b_{i\pm D}$ in the same readout group is given by

\begin{equation}
\label{s4cmb_MM}
L^{\rm{in}}_{i, i \pm D} = \frac{\alpha_{(i\pm D)}}{D^\beta}d^{(i \pm D)}\,\, .
\end{equation}
Assuming that the tones are linearly spaced within the frequency comb, we choose $\beta$ = 2. Additionally, we can turn on and off the intra-comb crosstalk, $L^{\rm{out}}_{ij}$, which is parameterized by a coefficient $\iota$ as
\begin{equation}
\label{Lout}
L^{\rm{out}}_{i,j}  = \iota \,\, ,
\end{equation}
where i and j are different SQUIDs and $\iota$ is fixed at 0.01\%.

\subsection{Systematic Effects From Crosstalk}\label{sec:systematics_xtalk}

In principle, crosstalk contamination can be corrected in data analysis by measuring the matrix elements of $L_{ij}$ using a number of analysis methods. Below, we assess the impact of crosstalk for simplified models of $L_{ij}$ assuming that no such mitigation can be performed in data analysis.

\subsubsection{Temperature to Polarization Leakage with Pair-Differencing}

In an experiment where the polarization signal on the sky is measured by differencing the signals from co-pointed detectors sensitive to orthogonal polarization states (pair-differencing), the primary impact of crosstalk is to leak a distorted copy of the temperature signal, either from crosstalk within the pixel pair or crosstalk from detectors of a different pixel pair, into polarization. In Sec.~\ref{sec:xtalk_hwp}, we will consider the case of experiments with a CRHWP, where the dominant effect is polarization to polarization leakage. It should be noted that many map-making algorithms do not explicitly difference the TOD from pairs of detectors, but the effect is expected to look similar in a maximum likelihood pipeline. We do not simulate this case due to the complexity and computational cost. 

To visualize this, consider a toy model of two pairs of detectors sensitive to orthogonal polarizations indexed $i,j$ and $k,l$ observing Stokes parameters \{$I, Q, U$\} with parallactic angle $\phi$ where subscript $1,2$ denotes different pointing.

\begin{gather}
d_i = I_1 + \frac{1}{2}\operatorname{Re}\{(Q_1 + iU_1) e^{-2i\phi_1}\} \,\, .\\
d_j = I_1 - \frac{1}{2}\operatorname{Re}\{(Q_1 + iU_1) e^{-2i\phi_1}\} \,\, .\\
d_k = I_2 + \frac{1}{2}\operatorname{Re}\{(Q_2 + iU_2) e^{-2i\phi_2}\} \,\, .\\
d_l = I_2 - \frac{1}{2}\operatorname{Re}\{(Q_2 + iU_2) e^{-2i\phi_2}\} \,\, .
\end{gather}

The polarization signal is formed by computing the difference:

\begin{equation}
d_i - d_j = \operatorname{Re}\{(Q_1 + iU_1) e^{-2i\phi_1}\} \, .\\
\end{equation}

The effect of crosstalk on this pair difference is:

\begin{equation}
\tilde{d}_i - \tilde{d}_j = \operatorname{Re}\{(Q_1 + iU_1) e^{-2i\phi_1}\} + (L_{ki} + L_{li} - L_{kj} - L_{lj})I_2 + (L_{ki} - L_{li} - L_{kj} + L_{lj})\operatorname{Re}\{(Q_2 + iU_2) e^{-2i\phi_2}\} \,\, .
\end{equation}

In the limit where $I\gg Q\simeq U$, which is motivated by the relative amplitudes of the CMB TT and EE spectra:

\begin{equation}
\tilde{d}_i - \tilde{d}_j \approx \operatorname{Re}\{(Q_1 + iU_1) e^{-2i\phi_1}\} + (L_{ki} + L_{li} - L_{kj} - L_{lj})I_2\,\,\,.
\end{equation}

Note that it is expected that the mixing matrix element $L_{ij}$ will be canceled by whatever calibration is used to match the gains of the two detectors in a pair. Crosstalk provides a means for the temperature signal in one detector to leak into the polarization channel of another detector pair. The exact magnitude of the resulting systematic contamination in polarization maps depends on the details of the matrix $L_{ij}$ and the observing strategy. 

\subsubsection{Polarization to Polarization Leakage with HWP Demodulation}\label{sec:xtalk_hwp}

In an experiment such as the SO SAT that employs a CRHWP to modulate the sky polarization signal, crosstalk does not leak temperature signal into the polarization channels. To visualize this, consider a data toy model for two detectors indexed \{$i,j$\} used in a HWP experiment (similar models can be found in \cite{Didier:EBEX_HWP}, \cite{Takakura:2017ddx}, or \cite{Kusaka:2013pla}):

\begin{gather}
d_i = I_1 + \operatorname{Re}\{(Q_1 + iU_1) e^{-2i\phi_i - 4i\chi}\} + A_i(\chi) \\
d_j = I_2 + \operatorname{Re}\{(Q_2 + iU_2) e^{-2i\phi_j - 4i\chi}\} + A_j(\chi) \,\, ,
\end{gather}
where $\chi = \omega_0 t$ is the HWP angle and $A(\chi)$ is a signal synchronous with the HWP angle. The crosstalk matrix $L_{ij}$ acts on these timestreams following

\begin{equation}
\tilde{d}_i = I_1 + L_{ji} I_2 + \operatorname{Re}\{[(Q_1 + iU_1) e^{-2i\phi_i} + L_{ji}(Q_2 + iU_2) e^{-2i\phi_j}] e^{-4i\chi}\} + A_i(\chi) + L_{ji} A_j(\chi)\,\, .
\end{equation}

In data analysis, the data is first treated by removing the signal (approximately) synchronous in $\chi$ either by subtracting a truncated Fourier series of the timestream binned in $\chi$ or by modulating the data by $e^{i n\chi}$ and subtracting a low order polynomial in time \cite{Takakura:2017ddx,Kusaka:2013pla}. Note that the cross-talked $A_j(\chi)$ is still synchronous with HWP angle and will be subtracted by either of these procedures, so this term is dropped.

\begin{equation}
\tilde{d}_i \rightarrow I_1 + L_{ji} I_2 + \operatorname{Re}\{[(Q_1 + iU_1) e^{-2i\phi_i} + L_{ji}(Q_2 + iU_2) e^{-2i\phi_j}] e^{-4i\chi}\}
\end{equation}

The central assumption of the HWP is that the input temperature signal $I$ has a sufficiently red spectrum and the HWP frequency $\omega_0$ is sufficiently fast that the polarization information can be separated in temporal frequency from the intensity information. To recover the temperature signal, a low pass filter is applied to the timestreams

\begin{gather}
\tilde{I}_1 \approx \textrm{LPF} \{I_1 + L_{ji} I_2 + \operatorname{Re}\{[(Q_1 + iU_1) e^{-2i\phi_i} + L_{ji}(Q_2 + iU_2) e^{-2i\phi_j}] e^{-4i\chi}\} \} \\
\tilde{I}_1 \approx I_1 + L_{ji} I_2\,\, .
\end{gather}

To recover the polarization signal, the detector timestream is multiplied by the conjugate of the modulation function $e^{-4i\chi}$ and low pass filtered:

\begin{equation}
(\tilde{Q}_1 + i\tilde{U}_1) e^{-2i\phi_i} \approx \textrm{LPF} \{e^{4i\chi}(I_1 + L_{ji} I_2) + e^{4i\chi} \operatorname{Re}\{[(Q_1 + iU_1) e^{-2i\phi_i} + L_{ji}(Q_2 + iU_2) e^{-2i\phi_j}] e^{-4i\chi}\} \}\,\, .
\end{equation}

Assuming the temperature signal is negligible in the polarization band,

\begin{equation}
(\tilde{Q}_1 + i\tilde{U}_1) e^{-2i\phi_i} \approx e^{4i\chi} \operatorname{Re}\{[(Q_1 + iU_1) e^{-2i\phi_i} + L_{ji}(Q_2 + iU_2) e^{-2i\phi_j}] e^{-4i\chi}\}\,\, . \\
\end{equation}

Expanding the exponentials into sines and cosines shows that crosstalk leaks polarization information from one channel to another with a phase set by the difference in detector polarization angles:

\begin{equation}
(\tilde{Q}_1 + i\tilde{U}_1) \approx (Q_1 + iU_1) + L_{ji}(Q_2 + iU_2) e^{-2i(\phi_j - \phi_i)} \,\, .
\end{equation}

This means that in a CRHWP experiment, detector crosstalk leaks temperature and polarization signal from one detector into another, but does not contaminate the polarization channels with the brighter temperature signal. The CRHWP is thus expected to significantly mitigate the contamination from detector crosstalk. The size of the contamination signal it maps depends on the exact details of $L_{ij}$ and the sky rotation achieved by the observation strategy.

\subsection{Crosstalk Simulation Results}\label{sec:xtalk_results}
In this section, we review the systematic effects induced on maps and their power spectra from the crosstalk added in \textit{s4cmb} as described in Sec.~\ref{Implementation}. We then offer design recommendations for the SO readout architecture based on these simulations.

\subsubsection{Power Spectrum Contamination}

We simulate the effect of instrumental crosstalk on the final power spectrum estimation using the \textit{s4cmb} framework. The simulation begins with $\Lambda$CDM skies containing only temperature and E-mode signals. These maps are then scanned to form TOD which are then perturbed by a crosstalk model and reprojected to the sky. The pointing used is ten days of a deep field scan targeting 1\% of the sky. The contamination is the power spectrum of the difference maps corrected for E-to-B leakage due to finite sky coverage \cite{2009PhRvD..79l3515G}. We run these simulations for both the pair-differencing and HWP cases.

\begin{figure}[h!]
\centering
\includegraphics[width=0.5\textwidth]{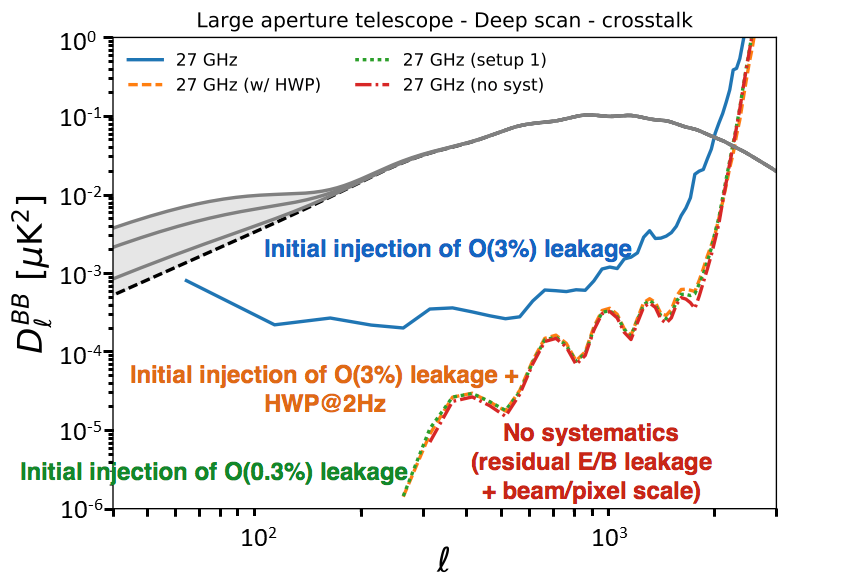}
\caption{Difference between the BB power spectrum of the LAT with crosstalk systematic and without ($C^{BB}_{\ell,\rm{sys}}$ - $C^{BB}_{\ell,\rm{no\: sys}}$), for the 27~GHz band with 3\% Lorentzian tail crosstalk with (solid blue) and without (dashed orange) a HWP, for 0.3\% without a HWP (dotted green), and with only pipeline residuals (red). The HWP drastically reduces the level of polarization leakage. Because the SO LAT will not use a HWP, the 27~GHz band crosstalk requirement will be more stringent. Simulations like this will be used to inform the hardware requirements for the LAT.}
\label{fig:Effect of HWP}
\end{figure}

\begin{figure}[h!]
\centering
\includegraphics[width=0.5\textwidth]{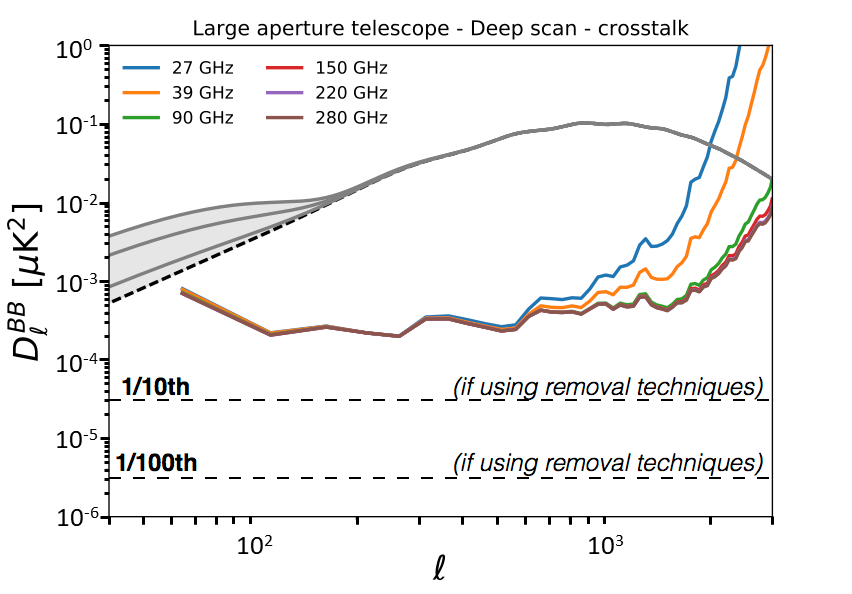}
\caption{Difference spectrum between the BB power spectrum with 3\% crosstalk systematic and without ($C^{BB}_{\ell,\rm{sys}}$ - $C^{BB}_{\ell,\rm{no\: sys}}$), for all of the approximate frequency bands in SO. The effect of crosstalk removal by measuring a mixing matrix is show in dotted lines. We expect this suppression to be $\geq 10$ times for the LAT based on recent results from SPTpol\cite{SPTpol2018}.}
\label{fig:XtalkVsBand}
\end{figure}

The power spectrum contamination is shown in Fig.~\ref{fig:Effect of HWP} for the 27~GHz band at the two average crosstalk levels studied, 3.0\% and 0.3\%. We consider the low frequency case because this is expected to have maximal contamination when the beam profile is included. This shows that in the pair-differenced case 3\% crosstalk without any mitigation produces a non-negligible bias at large angular scales, while the contamination due to 0.3\% crosstalk is significantly sub-dominant to the B-mode signal itself. The pipeline residuals are due to the imperfect separation of E- and B- modes on a pixelized sky and are also sub-dominant to the expected cosmological B-mode signal. A CRHWP does not mix temperature into polarization and suppresses the crosstalk effect. In this simulation the 3\% crosstalk becomes deeply sub-dominant to the B-mode signal itself. It should be noted that while a CRHWP will be used in the SATs, CRHWPs will not be used in the LAT. Figure~\ref{fig:XtalkVsBand} shows the added B-mode power without a HWP at a 3.0\% crosstalk level for the approximate SO bands. You can see that there is a stronger bias at $\ell \geqslant 1000$ for the lower bands (27 and 39~GHz) due to the beam size correction.

It is also possible to significantly suppress the effect of crosstalk in data analysis. This can be done by directly measuring the mixing matrix elements $L_{ij}$ and inverting the crosstalk in the time domain. In this case, the systematic error is not driven by the crosstalk matrix $L_{ij}$ but rather by the accuracy with which it can be measured by point source observations. We estimate this by showing the power spectrum assuming a factor 10 and a factor 100 suppression in power spectrum space. SPTpol has demonstrated a factor $\sim$ 10$\times$ suppression using this technique in an instrument with similar angular resolution to the LAT~\cite{SPTpol2018}.

We note that the presence of crosstalk results in a miscalibration of the instrument's beam profile by introducing negative sidelobes outside of a detector's main beam, which can cause additional power spectrum contamination that is not included in this work~\cite{Crites:2014prc}. These negative sidelobes are expected to differ between temperature and polarization for both pair-differencing and CRHWP experiments. We plan to study this effect in future work.

In the following sections, we will consider two crosstalk scenarios: crosstalk between two detectors in the same frequency band but at different spatial positions and crosstalk in different frequency bands but at the same spatial position.

\subsubsection{Crosstalk Within the Same Frequency Band}
The second crosstalk term we model, $L^{\rm{out}}_{ij}$, is constant for all bolometers on a readout comb. This component of crosstalk will only mix T$\rightarrow$T since both orthogonal bolometers in a pair will receive the same crosstalked signal, which will cancel out.

\begin{figure}[h!]
\centering
\includegraphics[width=0.9\textwidth]{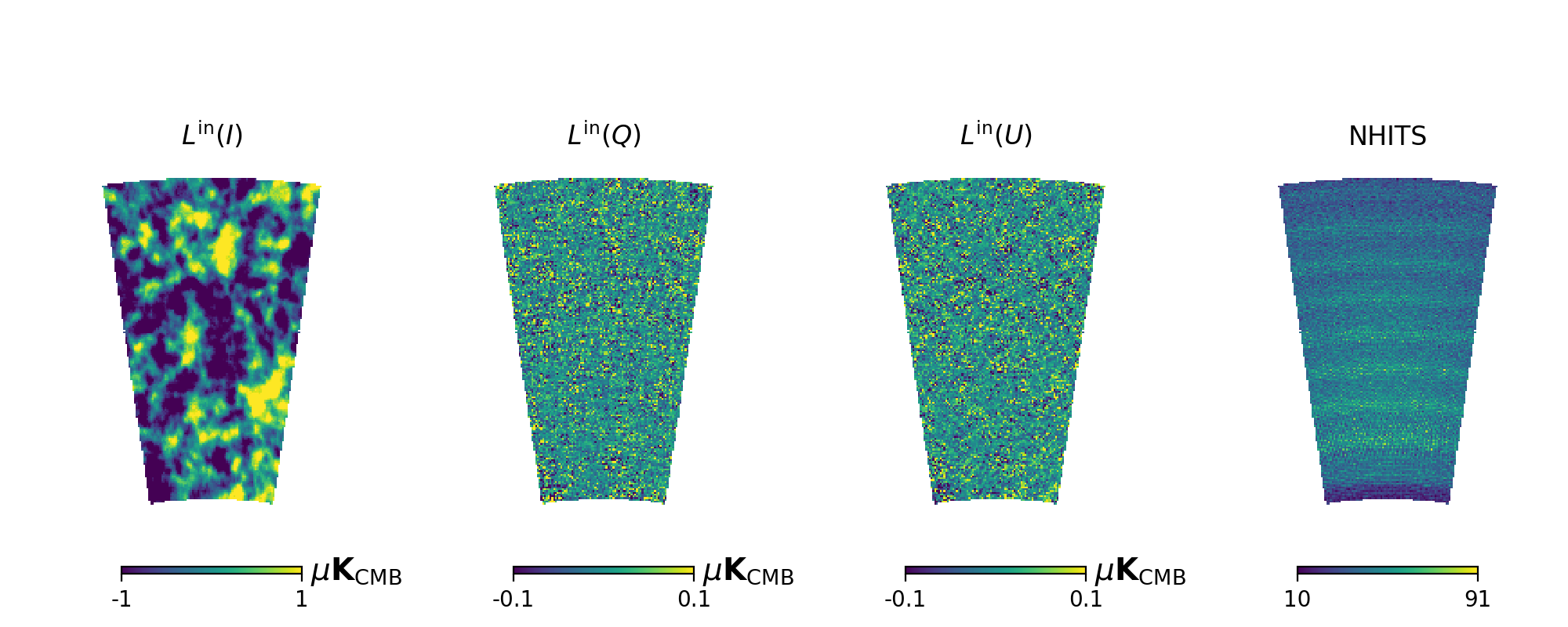}
\caption{Difference maps of one scan of the deep patch with and without Lorentzian tail crosstalk, $L^{\rm{in}}_{ij}$, added to the TOD show T$\rightarrow$T and T$\rightarrow$P leakage from mid to small angular scales.}
\label{fig:NN_Crosstalk}
\end{figure}

\begin{figure}[h!]
\centering
\includegraphics[width=0.9\textwidth]{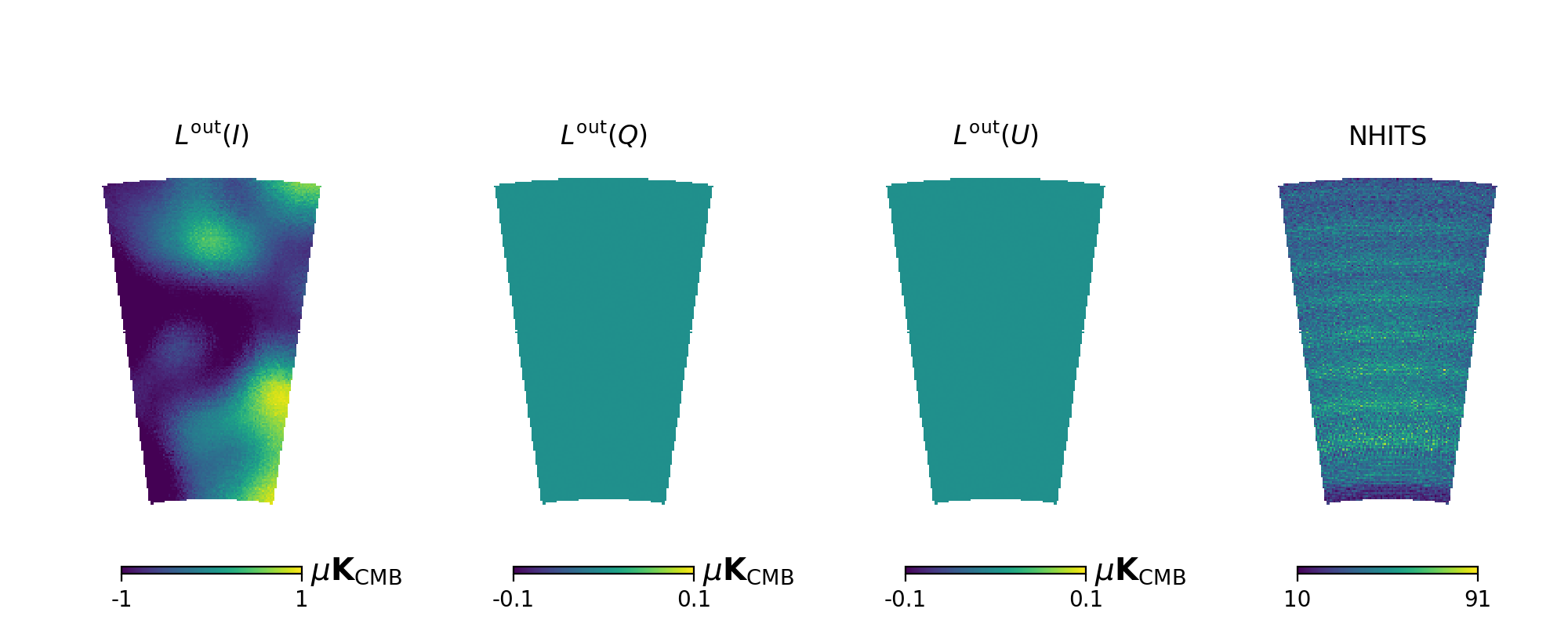}
\caption{Difference maps of one scan of the deep patch with and without inter-readout comb crosstalk, $L_{out}$, added to the TOD show T$\rightarrow$T leakage to large angular scales.}
\label{fig:LargeScaleLeakage}
\end{figure}

Crosstalk mixes angular scales on the sky. In a nearest-neighbor crosstalk model $L^{\rm{in}}_{ij}$, angular scales larger than the separation of pixels will be nearly unchanged and the predominant effect will be at scales equal to or less than the spacing between pixels. For crosstalk between detectors on different readout lines $L^{\rm{out}}_{ij}$ small and large angular scales will both be mixed. This effect can be seen in the difference maps shown in Figs.~\ref{fig:NN_Crosstalk} and~\ref{fig:LargeScaleLeakage} for $L^{\rm{in}}_{ij}$ and $L^{\rm{out}}_{ij}$, respectively. For clarity the maps shown are from a single one-hour observation. The effect will be suppressed due to sky rotation between different scans. Comparing the magnitude of the effect between the two models shows that $L^{\rm{out}}_{ij}$ must be $\sim$ 0.01\% to be on the same scale as that the $L^{\rm{in}}_{ij}$ model with $\alpha_{i\pm 1}\sim$ 1.0\%.


\subsubsection{Crosstalk Between Frequency Bands}

Crosstalk between detectors measuring different frequency bands has the potential to be a very difficult systematic to control. This is due to the fact that narrow-band sources do not exist on the sky. Point sources in the science or calibration data can be used to estimate the crosstalk matrix elements $L_{ij}$ for detectors at different positions; however, no such analog exists for inter-band crosstalk. We rely on the different spectral shapes of the CMB and our primary foregrounds (galactic dust and synchrotron radiation) as well as measurements at many different frequencies to produce a clean map of the CMB~\cite{2011PhRvD..84f9907E}. Inter-band crosstalk mixes different frequency maps causing a systematic bias in the separation of the CMB signal and foregrounds. This systematic must be adequately controlled in the design of the readout system since it is less straightforward to calibrate and remove in analysis. It is still possible to measure this mixing matrix $L_{ij}$ using a Fourier transform spectrometer, bias step ticking, or cosmic ray glitches.

\begin{figure}[h!]
\centering
\includegraphics[width=1.0\linewidth]{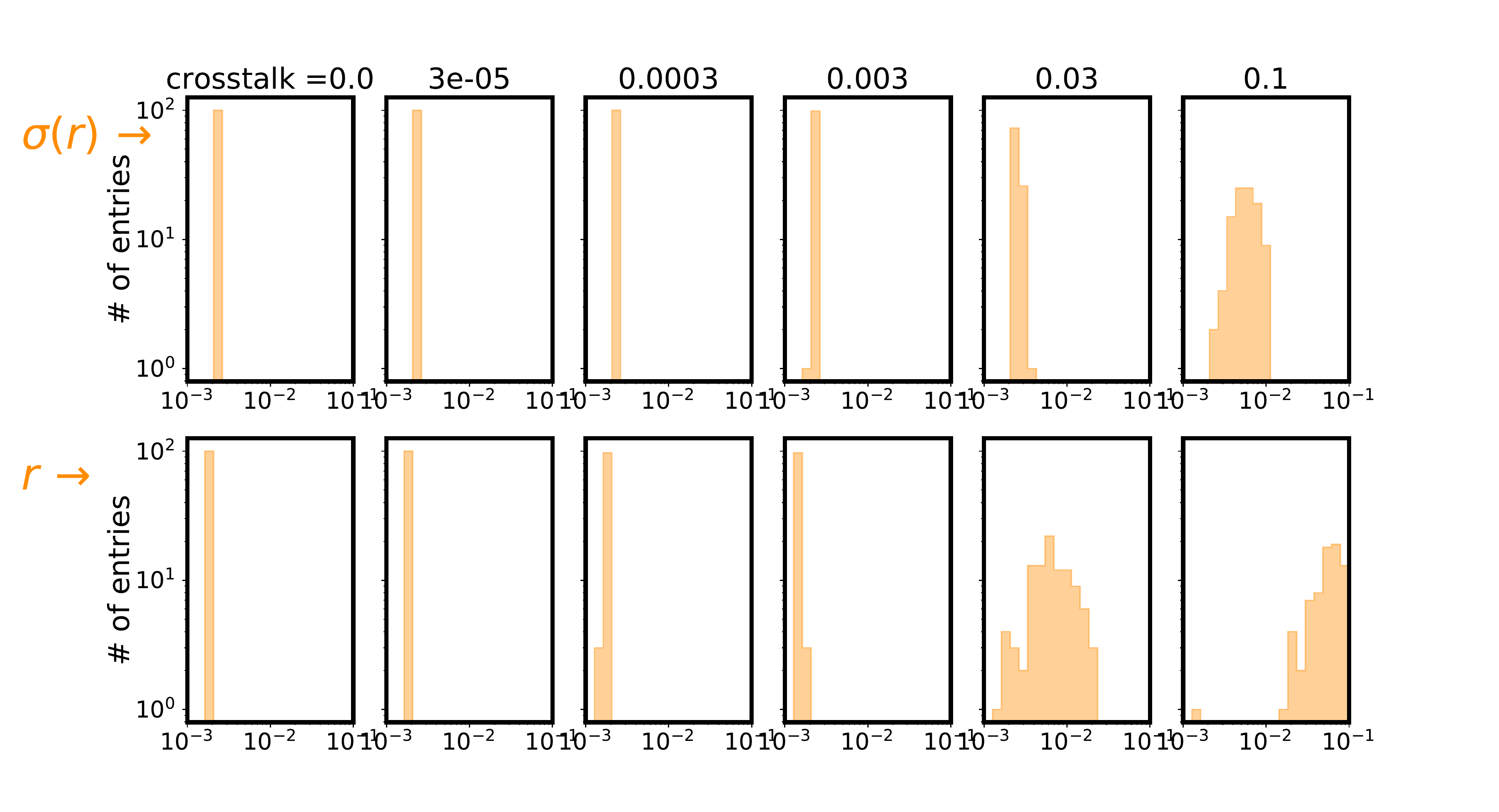}
\caption{Similar to the approach adopted in Ward, et al., 2018~\cite{Ward_2018}, we show the bias on the recovered tensor-to-scalar ratio $r$ (lower row) and the associated statistical uncertainty, $\sigma(r)$ (upper row), for different crosstalk amplitudes above. These results are derived using xForecast~\cite{Stompor_2016}, a component separation and forecast tool. Each column is estimated from 100~simulations of the crosstalk amplitudes drawn from a Gaussian probability distribution that is centered on zero with a width between 0 and $10\%$ as specified at the top of the panels. The case without crosstalk (leftmost column) shows a bias due to the spatial variability of foreground spectral indices which is not accounted for in the component separation analysis.}
\label{fig:IBXtalk}
\end{figure}

We use a simplified map-domain pipeline to directly mix the observed maps between frequencies and perform component separation on the distorted maps. This is equivalent to a model wherein there is only crosstalk between detectors measuring different frequencies at the same position and polarization angle. 
We exploit the xForecast~\cite{Stompor_2016} algorithm to perform CMB- and noise-averaged forecasts of the component separation, along with the estimation of the tensor-to-scalar ratio $r$. By assuming $r=0$ for the input sky maps (which are created using default PySM models~\cite{Thorne_2017}), the estimation of a non-zero tensor-to-scalar ratio after the foregrounds cleaning is the indication of a bias due to an imperfect recovery of the synchrotron or dust spectral energy distribution (SED). This can be due to both limitations in the algorithm itself (lack of internal degrees of freedom to model the complexity of the foregrounds emission, such as the spatial variability of SEDs) or due to systematic effects, such as crosstalk, in the input frequency maps.

The results of this study can be seen in Fig.~\ref{fig:IBXtalk}, where we depict the bias on $r$ with its associated error bar, $\sigma(r)$, after 100 simulations of crosstalk amplitudes drawn from Gaussian probability distribution. From these results, we can see that the residuals associated with inter-band crosstalk start to significantly impact the estimation of $r$ around and above $\sim$1\% crosstalk between any two frequency bands. The bias is not negligible at lower levels of crosstalk because of the complexity of foreground emissions. The spectral indices of synchrotron and dust are spatially varying in the input frequency map simulations, but we do not try to fit for this property in the component separation analysis. It is possible to further marginalize the likelihood on $r$ over some model of the foreground residuals, leading to a reduction of the bias but also a larger statistical error bar. Nevertheless, simulations with crosstalk amplitude larger than $1\%$ bias $r$ at the level of or above the statistical uncertainty.

It will be important to model this effect in our time-domain simulations. The scan strategy may significantly mitigate this effect, especially in the SAT where foregrounds will be a more significant challenge. These results can be taken as an upper bound for the design of the instrument.

\subsubsection{Informing the Design Process}

Several important conclusions can be drawn from these initial studies. Generally, it is very difficult to measure crosstalk between detectors in different frequency bands with data due to the lack of narrow-band sources on the sky. As a result, detectors measuring different frequency bands should be isolated in the readout as much as possible. Within the same frequency band it will be important to randomize the relative positions between nearest neighbors in readout frequency. Importantly, this point has recently been made in the context of Microwave Kinetic Inductance Detector resonators as well~\cite{MKID_Crosstalk}. This will take advantage of the natural parallactic angle rotation in a realistic scan strategy and detector multiplicity to average down the crosstalk. Additionally, the intra-band crosstalk requirements placed on the hardware will be significantly loosened in the SAT due to the mitigation provided by a continuously rotating HWP. 

\section{Conclusions and Future Work}
To fully leverage the sensitivity of SO, we must understand and control the level of systematics in the system. Modeling systematics in the design phase of the experiment provides critical feedback to the instrument design to ensure that SO meets its scientific goals. The preliminary studies on detector array systematic effects presented in this work have informed the spacing between pixels, detector parameter selection, the readout layout, and the scan strategy.

Studies of the optical coupling to the detector arrays show that the level of polarization leakage is at an acceptable level for the SO pixel sizes on the LAT and SAT with both architectures when beam calibration is included. Going forward, the map-based simulation pipeline developed for this study can be used to study the effect of any generic beam on the power spectra.

We have described the relevant models for long-timescale gain drifts that we expect to be sourced by our instrument's interactions with the TES bolometers in the focal plane. These effects are especially problematic for the LAT arrays. In the case of nonlinearity-induced differential gain between polarization-pair bolometers, the leakage of T$\rightarrow$P is not negligible given TES bolometer parameters optimized for background-limited performance and sensitivity on the LAT. However, we expect that, through more extensive future simulations, we can set acceptable upper bounds on these effects to avoid negative impacts on SO science goals by tuning relevant detector parameters. We additionally plan to constrain detector non-linearity through a direct measurement of higher-harmonic response, which will ensure that our TES bolometer arrays can perform as required. Further, we show how bath temperature fluctuations can couple into gain fluctuations of TES bolometers and discuss how this effect can also be controlled with bias power. Studies of both of these effects in the map and spectral domain are advancing toward SO-like detector counts and array configurations.

Crosstalk will be an important systematic to consider for the SAT and the LAT readout systems. The SAT will use a CRHWP that will significantly mitigate this effect. It will be important to simulate more realistic focal plane layouts and crosstalk matrix elements based on lab measurements. It will also be important to simulate the combined effect of crosstalk between detectors at different frequencies and spatial locations.

The simulations discussed in this work will inform hardware design choices and will incorporate more detail as the SO instrument designs mature. SO is a critical stepping stone for future experiments like CMB-S4. The tools and analyses developed for the SO systematic studies will be publicly released with the SO systematics studies to contribute to the design of future CMB experiments.

\acknowledgments      
 
This work was supported in part by a grant from the Simons Foundation (Award \#457687, B.K.). Some of the results in this paper have been derived using the HEALPix (Gorski et al., 2005)~\cite{Gorski:2005_healpix} package. 

\bibliographystyle{spiebib} 

\end{document}